\documentclass[preprint2]{aastex}

\usepackage{amsmath}
\usepackage{graphicx}
\usepackage{epsfig,psfrag,epic,eepic}
\usepackage{textcomp}
\usepackage{times}
\usepackage{longtable}

\slugcomment{Accepted for the Astrophysical Journal}

\shorttitle{Planets around massive stars}
\shortauthors{Janson et al.}

\begin{document}

\title{High-contrast Imaging Search for Planets and Brown Dwarfs around the Most Massive Stars in the Solar Neighborhood\altaffilmark{*}}

\author{Markus Janson\altaffilmark{1,6}, 
Mariangela Bonavita\altaffilmark{1}, 
Hubert Klahr\altaffilmark{2}, 
David Lafreni{\`e}re\altaffilmark{3}, 
Ray Jayawardhana\altaffilmark{1}, 
Hans Zinnecker\altaffilmark{4,5}
}

\altaffiltext{*}{Based on data from the Gemini North telescope under programs GN-2008B-Q-59, GN-2009A-DD-6, GN-2010A-Q-86 and GN-2010B-Q-14, on Keck data under program A200N2, and on Subaru data under program S09B-016.}
\altaffiltext{1}{Univ. of Toronto, 50 St George St, Toronto, ON M5S 3H8 Canada; \texttt{janson@astro.utoronto.ca}}
\altaffiltext{2}{Max Planck Institute for Astronomy, Heidelberg, Germany}
\altaffiltext{3}{University of Montreal, Montreal, Canada}
\altaffiltext{4}{Astrophysikalisches Institut Potsdam, Potsdam, Germany}
\altaffiltext{5}{SOFIA Science Center, NASA-Ames, Moffett Field, CA 94035, USA}
\altaffiltext{6}{Reinhardt fellow}

\begin{abstract}\noindent
There has been a long-standing discussion in the literature as to whether core accretion or disk instability is the dominant mode of planet formation. Over the last decade, several lines of evidence have been presented showing that core accretion is most likely the dominant mechanism for the close-in population of planets probed by radial velocity and transits. However, this does not by itself prove that core accretion is the dominant mode for the total planet population, since disk instability might conceivably produce and retain large numbers of planets in the far-out regions of the disk. If this is a relevant scenario, then the outer massive disks of B-stars should be among the best places for massive planets and brown dwarfs to form and reside. In this study, we present high-contrast imaging of 18 nearby massive stars, of which 15 are in the B2--A0 spectral type range and provide excellent sensitivity to wide companions. By comparing our sensitivities to model predictions of disk instability based on physical criteria for fragmentation and cooling, and using Monte-Carlo simulations for orbital distributions, we find that $\sim$85~\% of such companions should have been detected in our images on average. Given this high degree of completeness, stringent statistical limits can be set from the null-detection result, even with the limited sample size. We find that $<$32~\% of massive stars form and retain disk instability planets, brown dwarfs and very low-mass stars of $<$100~$M_{\rm jup}$ within 300~AU, at 99~\% confidence. These results, combined with previous findings in the literature, lead to the conclusion that core accretion is likely the dominant mode of planet formation.
\end{abstract}


\section{Introduction}

With the many hundreds of planet candidates from radial velocity \citep[e.g.][]{mayor2009,wright2011} and more than a thousand putative candidates from transits \citep[e.g.][]{carbera2009,borucki2011}, the population of close-in ($\sim$1~AU) exoplanets is starting to be reasonably well characterized down to sub-Jovian masses. Less is known about the wide ($>$10~AU) population of planets, but the range is starting to be probed with direct imaging. Several early surveys came up with null results \citep[e.g.][]{lafreniere2007b,kasper2007}, implying that very massive planets and brown dwarfs are rare in certain parameter ranges \citep{nielsen2010}. However, as instruments and techniques have improved, and as a wider range of target classes have been examined, substellar companions to stars have been increasingly discovered at a range of orbital separations \citep[e.g.][]{thalmann2009,biller2010,lafreniere2011}. In particular, some of the first probable giant planetary systems have been detected around A-stars \citep[e.g.][]{marois2008,lagrange2010}. While the close-in population of planets is believed to have formed by core accretion \citep[e.g.][]{mordasini2009}, the wide separations and high masses of some of these companions have led to the speculation that disk instabilities may play a role in the formation of such objects \citep[e.g.][]{boss2011}. The possible increase in frequency of massive planets around more massive stars implied by the direct imaging detections and shown more clearly in radial velocity surveys \citep[e.g.][]{hekker2008,johnson2010} may be attributed to the fact that more massive stars are expected to have more massive disks, which should benefit any planet formation mechanism. At some stellar mass, one might however expect the frequency of planets formed by core accretion to turn over, because the UV radiation increases with stellar mass, which decreases the primordial disk lifetime, and which may in turn impact core accretion since it requires formation timescales of a similar order, a few Myrs \citep[e.g.][]{alibert2005}. Disk instabilities, on the other hand, operate on much shorter timescales and should be insensitive to this effect. Hence, one might expect that the most massive stars should be the most beneficial hosts for the formation of massive giant planets and brown dwarfs through disk instability.

In this paper, we address the occurrence of massive planets and brown dwarfs around massive stars through high-contrast imaging of 18 of the heaviest stars in the solar neighborhood. We examine 15 stars in the B2--A0 spectral type range over almost the full parameter range in which disk instabilities could be expected to form substellar companions. We investigate two white dwarfs, likely remnants of very massive progenitors, over almost the full primordial range in which massive giant planets are expected to form or reside from core accretion. In addition, the very nearby A7-type star Altair is examined with a high sensitivity to giant planet companions. In the following, we first present the observations and data reduction, then the process of data analysis, comparison with theoretical models, and Monte Carlo simulations. This is followed by a discussion about core accretion and disc instability, and the implications of this study in that context. Finally, we make some concluding remarks.

\section{Observations and Data Reduction}

The observations presented in this paper were taken with adaptive optics using primarily NIRI \citep{hodapp2003} at Gemini North, and additionally IRCS \citep{kobayashi2000} at Subaru and NIRC2 at Keck. A summary of the observational parameters is provided in Table \ref{t:observations}. Our original plan for the NIRI data was to observe each star in both the CH4S- and Br$\alpha$-band. The CH4S-band then provides optimal sensitivity at large separations where the background noise sets the limit. At small separations, Br$\alpha$ observations would provide the best sensitivity, as the contrast performance in that band has been shown to be outstanding for bright stars with present-day instrumentation \citep{janson2008,janson2009}. However, during the first observations with NIRI+ALTAIR it became clear that the thermal background of this instrument combination is far higher than is the case for VLT/NACO (where the previous measurements had been made), hence the method could only provide useful information for the very brightest stars in the sample. Instead, we used the $K_{\rm c}$-band for small separations for the majority of the sample. The B2--A0-type stars in the sample were originally chosen based on the constraint that the brightness should be sufficient in $L^{\prime}$-band ($<$4 mag) to provide an adequate sensitivity in the Br$\alpha$-band, and based on their observability with the given instrument in the given observing period. Hence, the sample is brightness-limited (and thus largely volume-limited for a certain stellar mass). Some stars were removed from the list on the basis of binarity that would severely affect the achievable contrast or dynamical stability in the systems, but no obvious relevant bias results from this selection.

\begin{table*}[p]
\caption{Summary of observations}
\label{t:observations}
\centering

\begin{tabular}{lcccccccc}
\hline
\hline
Target & HIP ID & Instrument & Filter & Date & DIT (s) & Coadds & No. frames & Rotation (deg) \\
\hline
Alpheratz & 677 & NIRI & CH4S & 2008 Oct 14 & 30.0 & 1 & 90 & 64.9 \\
Alpheratz & 677 & NIRI & CH4S & 2009 Jul 30 & 30.0 & 1 & 83 & 93.0 \\
Alpheratz & 677 & NIRI & Br$\alpha$ & 2008 Sep 22 & 1.0 & 25 & 90 & 18.3 \\
41 Ari & 13209 & NIRI & Br$\alpha$ & 2008 Aug 29 & 1.0 & 25 & 90 & 83.5 \\
41 Ari & 13209 & NIRI & CH4S & 2008 Sep 24 & 30.0 & 1 & 90 & 74.4 \\
41 Ari & 13209 & NIRI & CH4S & 2010 Aug 23 & 30.0 & 1 & 90 & 64.4 \\
Algol & 14576 & NIRI & Br$\alpha$ & 2008 Sep 24 & 1.0 & 25 & 89 & 31.0 \\
Algol & 14576 & NIRI & CH4S & 2008 Oct 18 & 30.0 & 1 & 90 & 34.1 \\
Algol & 14576 & NIRI & CH4S & 2009 Sep 27 & 30.0 & 1 & 45 & 37.6 \\
Algol & 14576 & NIRI & $H$ & 2009 Sep 27 & 30.0 & 1 & 45 & 34.2 \\
Bellatrix & 25336 & NIRI & CH4S & 2008 Oct 18 & 30.0 & 1 & 90 & 46.9 \\
Bellatrix & 25336 & NIRI & $K_{\rm c}$ & 2008 Nov 25 & 30.0 & 1 & 68 & 33.3 \\
Bellatrix & 25336 & NIRI & CH4S & 2010 Oct 07 & 30.0 & 1 & 90 & 46.9 \\
Elnath & 25428 & NIRI & $K_{\rm c}$ & 2008 Nov 25 & 30.0 & 1 & 68 & 63.4 \\
Elnath & 25428 & NIRI & CH4S & 2008 Nov 30 & 30.0 & 1 & 90 & 40.9 \\
Elnath & 25428 & NIRI & CH4S & 2010 Oct 9 & 30.0 & 1 & 90 & 67.4 \\
LHS 1870 & 32560 & NIRC2 & $H$ & 2009 Dec 17 & 30.0 & 1 & 110 & 58.8 \\
$\beta$ CMi & 36188 & NIRI & $K_{\rm c}$ & 2008 Nov 5 & 30.0 & 1 & 68 & 39.2 \\
$\beta$ CMi & 36188 & NIRI & CH4S & 2008 Nov 14 & 30.0 & 1 & 90 & 56.6 \\
$\beta$ CMi & 36188 & NIRI & CH4S & 2010 Nov 28 & 30.0 & 1 & 90 & 43.0 \\
30 Mon & 41307 & NIRI & $K_{\rm c}$ & 2008 Nov 5 & 30.0 & 1 & 68 & 27.2 \\
30 Mon & 41307 & NIRI & CH4S & 2008 Nov 14 & 30.0 & 1 & 90 & 30.0 \\
30 Mon & 41307 & NIRI & CH4S & 2010 Nov 28 & 30.0 & 1 & 88 & 26.2 \\
$\theta$ Hya & 45336 & NIRI & $K_{\rm c}$ & 2008 Nov 30 & 30.0 & 1 & 68 & 35.2 \\
$\theta$ Hya & 45336 & NIRI & CH4S & 2008 Dec 4 & 30.0 & 1 & 90 & 40.5 \\
Regulus & 49669 & NIRI & CH4S & 2008 Dec 20 & 30.0 & 1 & 90 & 51.9 \\
Regulus & 49669 & NIRI & $K_{\rm c}$ & 2010 Feb 23 & 1.0 & 30 & 90 & 67.4 \\
GD 140 & 56662 & NIRC2 & $H$ & 2009 Dec 17 & 30.0 & 1 & 317 & 91.4 \\
$\gamma$ Crv & 59803 & NIRI & $K_{\rm c}$ & 2009 Jan 28 & 30.0 & 1 & 68 & 18.2 \\
109 Vir & 72220 & NIRI & CH4S & 2010 Jul 8 & 30.0 & 1 & 45 & 34.8 \\
109 Vir & 72220 & NIRI & CH4S & 2010 Jul 8 & 1.0 & 30 & 45 & 35.3 \\
$\beta$ Lib & 74785 & NIRI & CH4S & 2010 Jul 9 & 30.0 & 1 & 45 & 28.6 \\
$\beta$ Lib & 74785 & NIRI & CH4S & 2010 Jul 9 & 1.0 & 30 & 45 & 31.8 \\
$\epsilon$ Her & 83207 & NIRI & CH4S & 2010 Jun 29 & 30.0 & 1 & 45 & 38.9 \\
$\epsilon$ Her & 83207 & NIRI & CH4S & 2010 Jun 29 & 1.0 & 30 & 45 & 42.4 \\
Vega & 91262 & NIRI & Br$\alpha$ & 2008 Aug 31 & 1.0 & 25 & 189 & 71.6 \\
$\lambda$ Aql & 93805 & NIRI & CH4S & 2008 Aug 26 & 30.0 & 1 & 68 & 25.4 \\
$\lambda$ Aql & 93805 & NIRI & Br$\alpha$ & 2008 Sep 06 & 1.0 & 25 & 90 & 23.0 \\
$\lambda$ Aql & 93805 & NIRI & CH4S & 2010 Jun 16 & 30.0 & 1 & 90 & 28.1 \\
Altair & 97649 & IRCS & Br$\alpha$ & 2009 Aug 15 & 0.5 & 60 & 108 & 29.7 \\
\hline
\end{tabular}
\end{table*}

In all cases, the images were taken in pupil-stabilized mode rather than field-stabilized mode, to allow for high PSF stability and efficient implementation of angular differential imaging \citep[ADI, e.g.][]{marois2006}. The PSF core was allowed to saturate for the main ADI sequences of each target. For the NIRI Br$\alpha$- and $K_{\rm c}$-band data and the NIRC2 $H$-band data, short sequences with the star non-saturated were taken adjacent to the ADI observations for photometric calibration. For the NIRI $K_{\rm c}$-band cases where non-saturated images could not be acquired due to the brightness of the star (such as for Regulus), non-saturated images in the same band of other stars in the sample were used as photometric references. In the case of the IRCS data of Altair, we used non-saturated images of HR~8799 taken adjacent to the Altair observation as photometric standard images. For the NIRI CH4S-band images, a ghost feature could be used for photometric calibration as described in \citet{lafreniere2007b}. For each case where one or several companion candidates were found in the first epoch observation, a follow-up observation was executed 1-2 years afterward with the same settings (with two exceptions, see Sect. \ref{s:results}).

Reduction of the ADI data was performed with a LOCI-based \citep{lafreniere2007a} set of routines in IDL, as implemented in \citet{lafreniere2007b} for NIRI data and with customized adaptations for the other cameras. The LOCI parameters were chosen in the same way as in \citet{lafreniere2007b} for all the data, i.e. $N_{\rm \delta} = 0.5$, $N_{\rm A} = 300$, etc. Adaptations to other cameras were in the form of conforming to the structure of the data (different keywords and conventions), but the main LOCI parameters were the same, although in practice there are differences due to the fact that the typical FWHM in units of pixels for the respective instruments is different (3.5 pixels for NIRI in CH4S and $K_{\rm c}$, 4.0 pixels for NIRI in Br$\alpha$, 4.4 pixels for NIRC2 in $H$ and 6.3 pixels for IRCS in Br$\alpha$), which feeds into the size of the optimization regions and the selection of acceptable reference frames.

All the procedures include a correction for the partial subtraction imposed by the LOCI algorithm, as described in \citep{lafreniere2007a}, i.e. by introducing several fake companions into the images and running the pipeline on them to evaluate the flux loss. For the case of 4~$\mu$m data, we introduced an extra step analogous to that described in \citet{janson2008}, with unsharp-masking with a 500~mas gaussian kernel on the individual frames to remove low-frequency spatial variations in the thermal background.

\section{Data Analysis and Results}
\label{s:results}

\subsection{Identification of Companion Candidates}

\begin{figure}[ph]
\centering
\includegraphics[width=8cm]{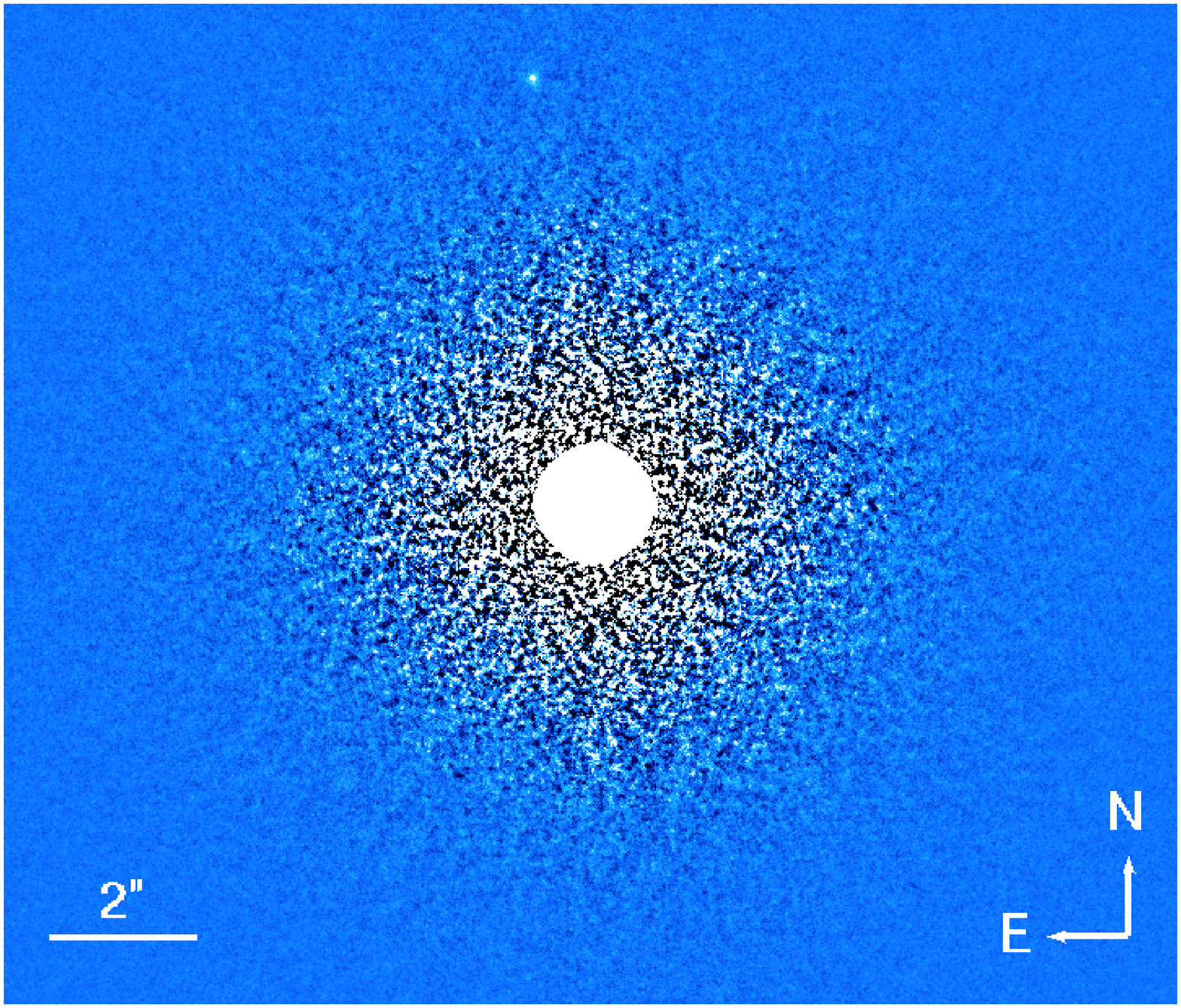}
\caption{\small{Example high-contrast image of the inner $\sim$7\arcsec radius region of Bellatrix, with a faint point-source to the North in the field of view. The source is 17.6 mag fainter than the star. A common-proper motion test has established that the candidate is likely a physcially unrelated background star (see Fig. \ref{f:ncpm}).}}
\label{f:25336}
\end{figure}

Each final reduced image (see Fig. \ref{f:25336} for an example) was carefully scanned for point sources, with visual inspection in the image itself, as well as in a corresponding $S/N$-map where sequential annuli of the image are normalized by the local noise in that annulus, and where a $S/N = 5$ detection criterion was used. For each candidate, we estimated astrometry and photometry relative to the primary. Astrometry was performed using Gaussian centroiding, and photometry using a circular aperture with a diameter of 77~mas, corresponding to a typical FWHM of the AO-corrected PSF in the 1.6~$\mu$m wavelength range. The astrometric error was assumed to be dominated by residual distortion correction errors as in e.g. \citet{lafreniere2011}, yielding errors of 6~mas in separation and 0.1~deg in position angle for the NIRI data. The photometric error was based on the local noise in an annulus at the separation of the point source. We list the properties of each point source in Appendix A. All brightness contrast values are provided in CH4S for the NIRI data, except for the case of $\gamma$ Crv which was only observed in $K_{\rm c}$. For the single point source detected in NIRC2 data, the contrast is in $H$. For this case, we determined astrometric errors using the pixel scale error of 0.006~mas/pixel and the orientation uncertainty of 0.02$^{\rm o}$ determined in previous astrometric calibrations using NIRC2 \citep{ghez2008}. No point sources were found in the IRCS Br$\alpha$ data, or indeed any of the Br$\alpha$ data from NIRI (one of the advantages of Br$\alpha$ is that background contamination is extremely low, since background stars are typically very faint with respect to the background at these wavelengths). 

\begin{figure}[ph]
\centering
\includegraphics[width=8cm]{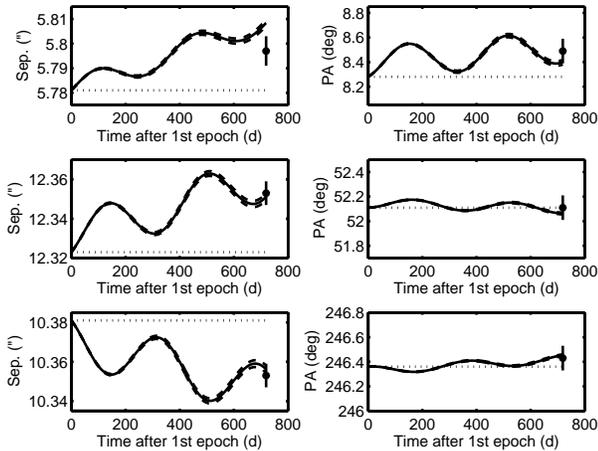}
\caption{\small{Proper motion tests for the three candidates in the field of Bellatrix. The upper, middle and lower rows are for the three different companions. The left column is for separation, and the right for position angle. The solid line in each panel is the expected motion of a background star, with the dashed lines corresponding to the errors. The dotted line marks common proper motion. The data point with errors is the measured position of the candidate at the second epoch measurement. All the candidates are probable background stars.}}
\label{f:ncpm}
\end{figure}

Although many of the stars had faint point sources in the field of view, none of these shared a common proper motion with the would-be host star, but were better consistent with the relative motion of a static background contaminant (with exceptions listed in the individual notes, including the case of Algol where additional observational criteria were applied for the companionship test). An example of a proper motion test is shown for Bellatrix in Fig. \ref{f:ncpm}, which is the "worst case" in the sense that it has by far the smallest proper motion between the two observational epochs, except for Algol. Two out of the three targets are fully consistent with the static background hypothesis. The remaining one is marginally deviant from the static background in separation (1.9$\sigma$), but it is closer to the background case than to the common-proper motion case, from which it deviates by 2.7$\sigma$. It also deviates from common-proper motion by 2.1$\sigma$ in position angle, where it is consistent with the static background. We therefore consider it as a likely background star, possibly with a non-zero proper motion of its own. All astrometric points for the candidates are listed in Appendix A. In summary, the survey revealed no substellar companions, and the data analysis was therefore focused on examining the implications of this absence of detections.

\subsection{Detection Limit Determination}

\begin{figure}[ph]
\centering
\includegraphics[width=8.0cm]{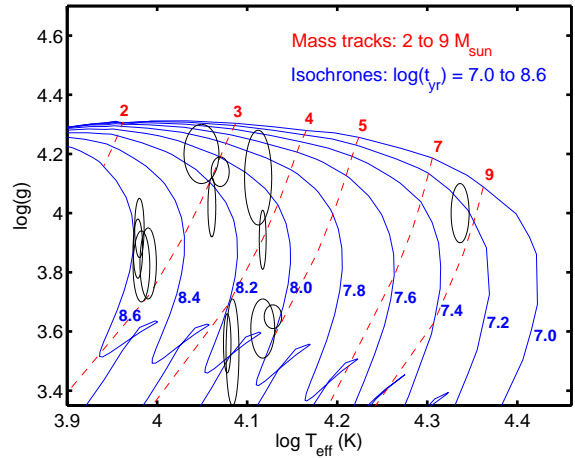}
\caption{\small{Isochrones and mass tracks from \citet{marigo2008} in $\log T_{\rm eff}$ versus $\log g$ for the B-star spectral type range. Dashed lines: Mass tracks for 2--9~$M_{\odot}$. Solid lines: Isochrones for ages of 10 to 400~Myr. Also plotted are the error ellipses for the 14 stars in the sample that were dated with the isochrone method. It can be seen that from $\sim$3~$M_{\odot}$ and upwards, the isochrones are sufficiently widely spaced with respect to the measurement errors that isochronal dating can be made with good precision.}}
\label{f:hrgt}
\end{figure}

The 5$\sigma$ brightness contrast limits as function of separation were evaluated by convolving each final image with a gaussian of the same FWHM as the observation, and subsequently taking the standard deviation within a sequence of annuli as the $\sigma$ for each corresponding separation. In order to determine what companions could have been detected, we use evolutionary models \citep{chabrier2000,baraffe2003} to translate brightnesses into masses. For this purpose, ages of the stars are required. Since B-stars evolve rapidly with time, isochronal dating yields their ages with good precision. Hence, most of the stars were dated with theoretical isochrones \citep{marigo2008} using values of $T_{\rm eff}$ and $\log g$ measured through spectroscopy from the literature \citep{adelman2000,gray2003,levenhagen2006,soubiran2010,wu2011}. In some cases this method was not applicable, these are noted in the individual notes of Sect. \ref{s:bstars} for the B2--A0 stars, and in Sect. \ref{s:wds} for LHS~1870 and GD~140, and Sect. \ref{s:altair} for Altair. The isochrones are shown in Fig. \ref{f:hrgt}, and the ages are listed in Table \ref{t:targets}. In general, the targets are only consistent with one or two isochrones, yielding uncertainties of 10--20~\%. Two exceptions are noted in the individual notes.

\begin{table*}[p]
\caption{Target properties}
\label{t:targets}
\centering

\small{
\begin{tabular}{lccccccccccc}
\hline
\hline
Target & HIP ID & RA & Dec & SpT & $H$\tablenotemark{a} & Dist.\tablenotemark{b} & $\mu_{\alpha}$ & $\mu_{\delta}$ & Age & Mass\tablenotemark{c} & [Fe/H]\tablenotemark{d} \\
 & & & & & (mag) & (pc) & (mas/yr) & (mas/yr) & (Myr) & ($M_{\odot}$) & (dex) \\
\hline
Alpheratz & 677 & 00 08 23.26 & +29 05 25.6 & B9II & 2.29 & 29.8 & 135.68 & -162.95 & $130^{+10}_{-20}$ & $4.0^{+0.2}_{-0.1}$ & 0.00  \\
41 Ari & 13209 & 02 49 59.03 & +27 15 37.8 & B8V & 3.80 & 48.9 & 65.47 & -116.59 & $130^{+10}_{-30}$ & $3.1^{+0.1}_{-0.1}$ & ...  \\
Algol & 14576 & 03 08 10.132 & +40 57 20.3 & B8V & 1.95 & 28.5 & 2.39 & -1.44 & $460^{+20}_{-20}$ & $2.8^{+0.5}_{-0.5}$ & ...  \\
Bellatrix & 25336 & 05 25 07.86 & +06 20 58.9 & B2III & 2.36 & 74.5 & -8.75 & -13.28 & $20^{+2}_{-4}$ & $8.4^{+0.3}_{-0.1}$ & -0.25  \\
Elnath & 25428 & 05 26 17.51 & +28 36 26.8 & B7III & 1.88 & 40.2 & 23.28 & -174.22 & $100^{+10}_{-10}$ & $5.0^{+0.1}_{-0.1}$ & -0.10  \\
LHS 1870 & 32560 & 06 47 37.99 & +37 30 57.1 & DA2 & 12.66 & 15.4 & -225.63 & -935.45 & $700^{+300}_{-200}$ & $3.1^{+0.5}_{-0.5}$ & ...  \\
$\beta$ CMi & 36188 & 07 27 09.04 & +28 36 26.8 & B8V & 3.11 & 52.2 & -50.28 & -38.45 & $160^{+20}_{-60}$ & $4.2^{+0.8}_{-0.1}$ & ...  \\
30 Mon & 41307 & 08 25 39.63 & -03 54 23.1 & A0V & 4.09 & 38.3 & -66.05 & -24.20 & $400^{+50}_{-40}$ & $2.5^{+0.1}_{-0.1}$ & -0.44  \\
$\theta$ Hya & 45336 & 09 14 21.86 & +02 18 51.4 & B9.5V & 4.04 & 39.5 & 112.57 & -306.07 & $130^{+30}_{-70}$ & $2.7^{+0.3}_{-0.1}$ & 0.40  \\
Regulus & 49669 & 10 08 22.31 & +11 58 01.9 & B7V & 1.66 & 23.8 & -249.40 & 4.91 & $160^{+20}_{-20}$ & $4.2^{+0.1}_{-0.1}$ & 0.00  \\
GD 140 & 56662 & 11 37 05.10 & +29 47 58.3 & DA3 & 13.11 & 15.4 & -146.90 & -5.91 & $250^{+30}_{-20}$ & $6.3^{+0.7}_{-0.7}$ & ...  \\
$\gamma$ Crv & 59803 & 12 15 48.37 & -17 32 30.9 & B8III & 2.83 & 50.6 & -159.58 & 22.31 & $160^{+40}_{-30}$ & $4.2^{+0.4}_{-0.3}$ & ...  \\
109 Vir & 72220 & 14 46 14.92 & +01 53 34.4 & A0V & 3.63 & 39.4 & -116.04 & -21.75 & $320^{+40}_{-40}$ & $3.0^{+0.1}_{-0.1}$ & -0.41  \\
$\beta$ Lib & 74785 & 15 17 00.41 & -09 22 58.5 & B8V & 2.89 & 49.1 & -96.39 & -20.76 & $80^{+50}_{-40}$ & $3.5^{+0.3}_{-0.2}$ & 0.33  \\
$\epsilon$ Her & 83207 & 17 00 17.37 & +30 55 35.1 & A0V & 3.64 & 49.9 & -47.68 & 26.89 & $400^{+50}_{-40}$ & $2.6^{+0.1}_{-0.1}$ & -0.25  \\
Vega & 91262 & 18 36 56.34 & +38 47 01.3 & A0V & -0.03 & 7.8 & 201.03 & 287.47 & $400^{+50}_{-40}$ & $2.5^{+0.1}_{-0.1}$ & -0.43  \\
$\lambda$ Aql & 93805 & 19 06 14.94 & -04 52 57.2 & B9V & 3.48 & 38.4 & -19.68 & -90.37 & $160^{+20}_{-20}$ & $3.1^{+0.1}_{-0.1}$ & 0.00  \\
Altair & 97649 & 19 50 47.00 & +08 52 06.0 & A7V & 0.10 & 5.1 & 536.87 & 385.57 & $500^{+250}_{-250}$ & $1.7^{+0.1}_{-0.1}$ & -0.24  \\
\hline
\end{tabular}
}
\tablenotetext{a}{From 2MASS \citep{skrutskie2006}. Typical uncertainties are $\pm$0.2 mag.}
\tablenotetext{b}{From Hipparcos \citep{perryman1997}. Typical uncertainties are 1--5~\% in distance and 0.5--1~mas/yr in proper motion.}
\tablenotetext{c}{Mass of the primary in the case of binaries, primordial mass for white dwarfs.}
\tablenotetext{d}{From \citet{gray2003,valdes2004,wu2011}. Typical uncertainties are 0.1 dex when quoted in the literature.}
\end{table*}

Since the evolutionary models only provide values in broad-band filters by default, we used a grid of COND spectra \citep{allard2001} for temperatures of 100~K to 1700~K and DUSTY spectra \citep{chabrier2000} for 1800~K to 3000~K (surface gravities of 3.5--5.0~dex in both cases) to re-calculate the evolutionary model values from broad-band values (e.g. $H$) to corresponding narrow-band values (e.g. CH4S) as function of age and companion mass.

\begin{figure}[ph]
\centering
\includegraphics[width=8cm]{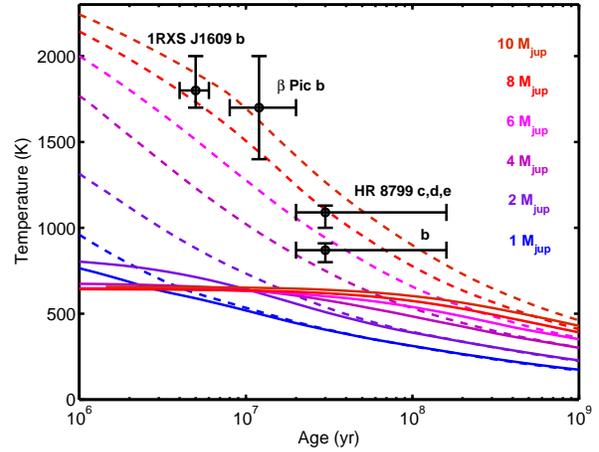}
\caption{\small{Temperature as a function of age for young planets, as predicted by the cold-start (solid lines) and hot-start (dashed lines) models from \citet{fortney2008}. Also plotted are the values for HR~8799~b--e \citep{marois2008,marois2010}, $\beta$~Pic~b \citep{bonnefoy2011}, and 1RXS~J1609~b \citep{lafreniere2008}. Although these objects may have formed by different mechanisms, they are all better described by hot-start than by cold-start models.}}
\label{f:hotcold}
\end{figure}

A common discussion regarding model inferences of parameters in direct imaging planet searches is the discussion of hot-start \citep[e.g.][]{baraffe2003,burrows2003} versus cold-start \citep[e.g.][]{marley2007,fortney2008} models. Cold-start models were introduced in an effort to represent core accretion formation of exoplanets, where the accretion shock may lead to signficantly cooler initial temperatures than in the hot-start representation of a gravitational collapse, although there is significant uncertainty about the magnitude of this effect, due to the 1D nature and simplified shock treatment of the cold-start models \citep{fortney2008}. For most of the discussion of this paper, this issue is irrelevant, since we are considering gravitional instability as formation mechanism, rather than core accretion. The exception is in the discussion of the white dwarfs and Altair, where formation by core accretion may be relevant. However, these systems are old enough and have low enough minimum detectable masses that the hot- and cold-start models will have largely converged. Furthermore, it is important to note that the cold-start models are not supported by observations, even in cases where core accretion seems to be the relevant mechanism. This is summarized in Fig. \ref{f:hotcold}, where temperature is plotted as function of age for directly imaged planets with constrained temperatures and small mass ratios relative to the parent star, along with cold- and hot-start models from \citet{fortney2008}. The hot-start models provide much better matches to the data points, and the cold-start models never reach the relevant temperatures, regardless of age. The most interesting case is that of $\beta$ Pic b, which is probably the strongest single case for having formed by core accretion, given its semi-major axis of only $\sim$10 AU. Hence, we do not consider cold-start conditions as relevant for this study. Nonetheless, it is relevant to point out that, as in any direct imaging effort of very low-mass companions, the inference of physical properties from the observational data depends on models that have not been calibrated against objects with known dynamical masses, thus some caution about their accuracy is justified.

The further data analysis is described separately for the cases of the 15 B2--A0 stars, the two white dwarfs, and Altair, respectively.

\subsection{The B2--A0 Sample}
\label{s:bstars}

The 15 stars in the B2--A0 sample are used to specifically address formation and in-situ retention of companions by disk instabilities, since the data provide very high sensitivity in the ranges where companions could be expected to form through such a mechanism. For this purpose, we use a model that predicts within which boundaries in mass/semi-major axis space fragments can form from instabilities (Klahr et al. 2011, in prep.). The principle behind the model is described in e.g. \citet{mordasini2011}, here we provide a brief summary. The model requires that for a fragment in the planet or brown dwarf mass range to form, at least two conditions need to be simultaneously fulfilled \citep{rafikov2005}. One condition is that the local surface density in the disk needs to be high enough for an azimuthally symmetric instability to form at a certain disk radius. This can be formulated as: 

\begin{equation}
Q = \frac{c_{\rm s} \kappa}{\pi G \Sigma} < 1
\end{equation}

where $Q$ is the Toomre parameter, $c_{\rm s}$ is the sound speed, $\kappa$ is the epicyclic frequency, and $\Sigma$ is the gas surface density \citep{toomre1981}. In the following, this will be referred to as the Toomre criterion. Another condition is that for a fragment to form out of the instability, the cooling time $\tau_{\rm cool}$ needs to be shorter than the shearing time \citep{gammie2001}, which in turn is of the same order as the orbital timescale $\tau_{\rm orb}$, i.e.:

\begin{equation}
\tau_{\rm cool} < \tau_{\rm orb}
\end{equation}

which will be referred to here as the cooling criterion. These conditions taken together mean that in a planetary mass versus semi-major axis diagram, there is a limited range where planets could form through gravitational instabilities. The boundaries of this area depend on the properties of the parent star, but as a general principle, the Toomre criterion states that instability occurs more easily for a higher local mass, so at a certain semi-major axis, only planets above a certain mass can form. Simultaneously, the cooling criterion states that fragmentation occurs more easily for longer orbital periods, so for a given planetary mass, only planets beyond a certain semi-major axis can form.

The model calculates the behaviour of fragmentation based on a 1D vertical disk structure code first developed in \citet{bell1997}, but with the additional inclusion of realistic radiation input from the central star, as in \citet{dalessio1999}. The cooling time is determined by first calculating the structure for the isothermal equilibrium case, where the irradiation temperature $T_{\rm irr}$, the effective temperature $T_{\rm eff}$ and the central temperature $T_{\rm cen}$ of the region are all equal. This includes a calculation of the thermal energy for this case, denoted $E_{\rm th,0}$. Following this procedure, a second structure is calculated for a given $T_{\rm cen} > T_{\rm eff}$ with a thermal energy denoted as $E_{\rm th,1}$. The cooling time can then be calculated from a comparison of these cases, using the proportionality:

\begin{equation}
\tau_{\rm cool} \propto \frac{E_{\rm th,1}-E_{\rm th,0}}{T_{\rm eff}^4-T_{\rm irr}^4}
\end{equation}

\begin{figure}[ph]
\centering
\includegraphics[width=8cm]{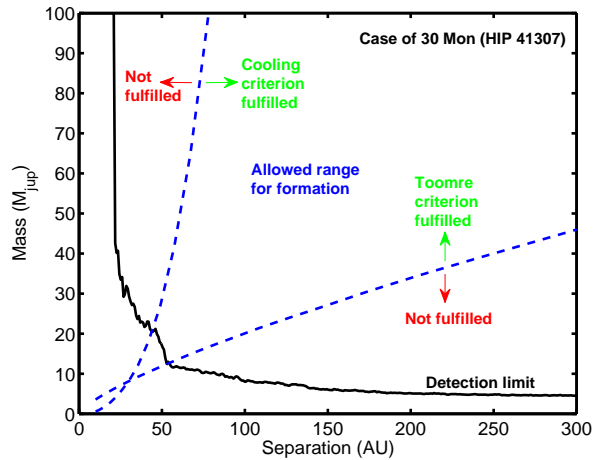}
\caption{\small{Example of model predictions and detection limit performance, for the case of 30~Mon. The dashed lines delimit the allowed range for formation, and correspond to the Toomre criterion and the cooling criterion, respectively. Only objects above the Toomre line and to the right of the cooling line are allowed. The solid line is the combined mass detection limit in  projected separation for comparison.}}
\label{f:41307}
\end{figure}

\begin{figure}[ph]
\centering
\includegraphics[width=8cm]{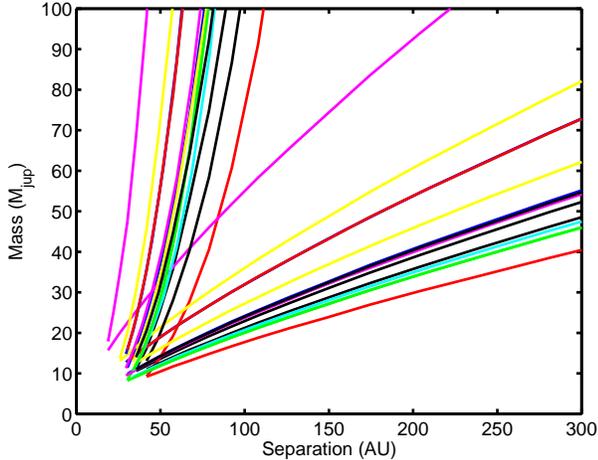}
\caption{\small{Limits from the Toomre criterion and cooling criterion beyond the crossover point for all the 15 stars in the B2--A0 sample, to illustrate the distribution of allowed ranges.}}
\label{f:toomre}
\end{figure}

\begin{figure}[ph]
\centering
\includegraphics[width=8cm]{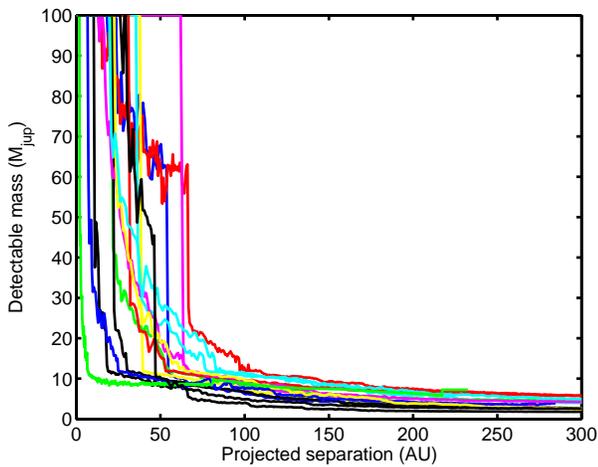}
\caption{\small{Combined mass detection limits for all the 15 stars in the B2--A0 sample, to illustrate the distribution of limits.}}
\label{f:masscurves}
\end{figure}

The most unstable wavelength \citep[e.g.][]{dangelo2010} is used for the purpose of converting disk gas surface density $\Sigma$ into the mass of the fragment. Thus, the model provides individual boundaries for each of the stars in the B2--A0 sample, taking the stellar mass, initial luminosity\footnote{In our case determined on the basis of isochrones from \citet{marigo2008}, at the zero-age main sequence.} and metallicity as input. Metallicity is adopted from spectroscopic results in the literature \citep{gray2003,valdes2004,wu2011}. In the few cases where no literature value exist (empty fields in Table \ref{t:targets}), Solar abundance is assumed. A possible caveat in this procedure is if planets form while the star is still accreting its envelope, in which case the mass and luminosity would be smaller. However, mass and luminosity have opposite effects on the boundaries, in the sense that a smaller luminosity lowers the Toomre criterion and pushes the cooling criterion outwards, while a smaller stellar mass raises the Toomre criterion and pushes the cooling criterion outward. Thus, the net effect is typically modest \citep[compare, e.g., with the case of the Sun-like star GJ~758 in][]{mordasini2011}. In Fig. \ref{f:41307} we show an example of boundaries for allowed planet formation by gravitational instability, for the case of 30~Mon. The mass sensitivity as a function of projected separation is also shown for comparison. This latter curve is a combination of the best detection region from each of the two observations performed for this target (one in CH4S and one in $K_{\rm c}$). To show the distribution of allowed formation regions for the full sample, we plot all of them in Fig. \ref{f:toomre}. Similarly, we show all the mass detection limits in Fig. \ref{f:masscurves}. It can be readily seen that the detection space covers a very large fraction of the formation space, although since the detection space is presented in projected separation, the comparison does not provide the full story. For that purpose, we use Monte Carlo simulations of orbital distributions, as described below.

In general, binaries with orbits that may cause instability or inhibit contrast performance in our detectable image space (separations of $\sim$0.3--10\arcsec) were removed from our input sample. However, there are two remaining targets in the sample that turn out to have such binary components ($\gamma$ Crv and 109 Vir, see individual notes). For these two cases, we impose the additional criterion that only companions with dynamically stable orbits can form in the system, where stable orbits are determined following the procedure in \citet{holman1999}. Since the orbits of the two binaries are unknown, we assume a semi-major axis of 1.31 times the projected separation, as in \citet{fischer2002} and an eccentricity of 0.36 \citep[the mean of the distribution in][]{duquennoy1991} in both cases. From this procedure, it follows that companions can only form outside of 241 AU for $\gamma$ Crv, and outside of 102 AU for 109 Vir.

Once the boundaries are calculated, we make two different lines of assumption for the distribution of where objects are formed and retained within these boundaries. One path is to assume that the distribution of objects is uniform over the allowed area. In terms of comparing to what we could detect, this is probably a conservative assumption, because the model refers to the formation of the initial fragment, which may subsequently accumulate further gas and increase in mass \citep[e.g.][]{kratter2010}, which would always make it easier to detect in the images. We will refer to this as the uniform case. The other path is yet more conservative, and assumes that the distribution is uniform along the minimum mass boundary set by the Toomre criterion. This corresponds to the extreme case in which the disk collapses entirely into minimum mass fragments, leaving no free gas to accumulate further. This case will be referred to as the minimum case. In addition to the boundaries from the above description, we also set an upper mass limit of 100~$M_{\rm jup}$ since we are only interested in very low-mass companions (in other words, our study does not exclude or otherwise address binary formation through disk instabilities, see Sect. \ref{s:discussion}). We also assume an outer disk radius of 300~AU. The outer radii of disks are typically poorly constrained, but millimeter interferometry observations have implied radii of a few to several hundred AU in the best studied cases \citep[e.g.][]{pietu2007}. Of course, there is a whole range of different types of assumptions that could be made in addition to those we have made here. For instance, one might imagine that the companion distribution could be uniform in logarithm space rather than linear space, which would give some additional weight to small separations which would affect the detection limit of the more distant of the targets. However, we restrict our analysis to the uniform and minimum cases described above, as the main examples.

The detectability of this artificial planet population is evaluted through use of the Monte-Carlo code MESS \citep{bonavita2011}. The code generates $10^4$ orbits per grid point in a mass versus semi-major axis grid with a sampling of 5~AU in semi-major axis and 1~$M_{\rm jup}$ in mass, for all grid points that fall into the allowed formation range for gravitational instability. The orbits are randomly oriented in space, and the orbital phase is also random. Since the eccentricity distribution is unknown, we made two separate simulations in which the orbits were purely circular in one case (referred to as the circular case), and where the eccentricities were randomly generated with a uniform distribution between 0.0 and 0.6 in the other case (referred to as the eccentric case). For each random event, the code then generates the projected separation. The mass of the artificial planet is compared to the detectable mass at that projected separation from the mass detection limits, and the fraction of orbits at each grid point that turn out to be detectable then corresponds to the probability of detection at that grid point. 

\begin{figure}[ph]
\centering
\includegraphics[width=8cm]{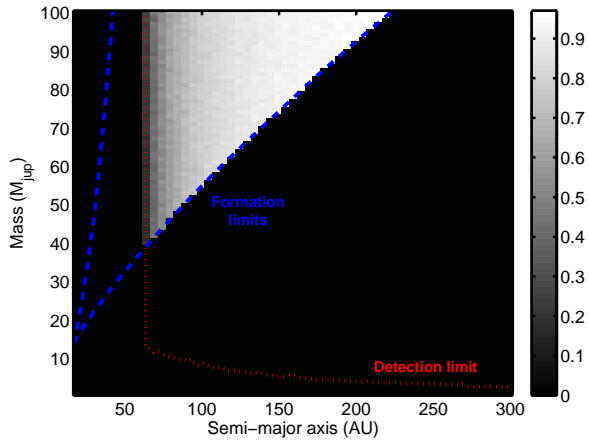}
\caption{\small{Detection probability map for Bellatrix, for the circular case.}}
\label{f:bellatrix_c}
\end{figure}

\begin{figure}[ph]
\centering
\includegraphics[width=8cm]{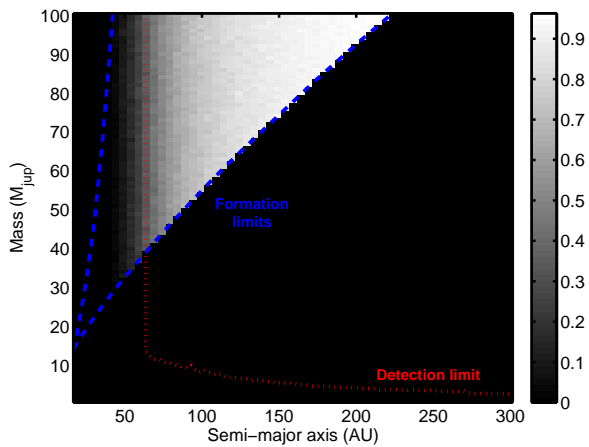}
\caption{\small{Detection probability map for Bellatrix, for the eccentric case.}}
\label{f:bellatrix_e}
\end{figure}

As can be expected from the fact that the mass detection limits cover such a large portion of the allowed formation space, it is indeed the case that the probabilities of detection evaluated over the full range are very high. It makes virtually no difference whether the orbits are circular or eccentric, and the difference between the uniform and minimum formation cases is also quite small. Average detection probabilities for the individual stars are listed in table \ref{t:detprob}. Typical probabilities are 80--90~\%. The only target with significantly less than 80~\% consistently for all cases is Bellatrix (see Fig. \ref{f:bellatrix_c} for the circular case and \ref{f:bellatrix_e} for the eccentric case), where almost half of the artifical population is undetectable. The reason for this is that this star has the largest distance in the sample, in combination with the fact that it has a low metallicity, which forces companion formation to relatively small separations. For similar reasons, Bellatrix has among the highest differences in detection probability between the circular and eccentric cases -- this is because it is the case where the detection limit has the largest impact on the observability of the population. The reason that the probability is higher in the eccentric case for Bellatrix is that eccentric orbits with semi-major axes just below the inner working angle of the image spend most of their time at apastron, thus providing a net benefit in observability relative to the circular case. Other cases that have relatively big differences between the circular and eccentric cases are Regulus and Vega, where the outer working angle is small enough that eccentric orbits at apastron near the outer edge of the field are preferentially lost relative to the circular case.

\begin{table*}[ph]
\caption{\small{Detection probabilities}}
\label{t:detprob}
\centering

\begin{tabular}{lcccc}
\hline
\hline
Target & Det. prob. & Det. prob. & Det. prob. & Det. prob. \\
 & circular & eccentric & circular & eccentric \\
 & uniform & uniform & minimum & minimum \\
\hline
Alpheratz	&	90.6~\%	&	88.8~\%	&	77.4~\%	&	76.1~\%	\\
41~Ari  	&	90.3~\%	&	90.0~\%	&	80.6~\%	&	80.8~\%	\\
Algol   	&	91.3~\%	&	87.7~\%	&	71.5~\%	&	69.2~\%	\\
Bellatrix	&	52.6~\%	&	54.3~\%	&	64.5~\%	&	64.7~\%	\\
Elnath  	&	90.8~\%	&	89.9~\%	&	87.4~\%	&	87.4~\%	\\
$\beta$~CMi	&	92.6~\%	&	91.9~\%	&	84.0~\%	&	84.2~\%	\\
30~Mon  	&	97.0~\%	&	96.9~\%	&	84.5~\%	&	84.0~\%	\\
$\theta$~Hya	&	98.6~\%	&	98.4~\%	&	96.7~\%	&	96.0~\%	\\
Regulus 	&	91.1~\%	&	82.7~\%	&	89.8~\%	&	80.7~\%	\\
$\gamma$~Crv	&	99.3~\%	&	99.2~\%	&	99.4~\%	&	99.3~\%	\\
109~Vir 	&	99.0~\%	&	98.8~\%	&	97.5~\%	&	97.3~\%	\\
$\beta$~Lib	&	96.3~\%	&	96.1~\%	&	86.0~\%	&	86.4~\%	\\
$\epsilon$~Her	&	94.6~\%	&	94.5~\%	&	75.3~\%	&	76.0~\%	\\
Vega    	&	86.8~\%	&	79.7~\%	&	82.6~\%	&	75.1~\%	\\
$\lambda$~Aql	&	94.1~\%	&	93.6~\%	&	86.4~\%	&	86.5~\%	\\
\hline
Sample mean & 91.0~\% & 89.5~\% & 84.2~\% & 82.9~\% \\
\hline
\end{tabular}
\end{table*}

Given these very high detection probabilities even after accounting for projection effects, some rather strong statistical constraints can be set on the companion frequency from disk instability, despite the limited size of the sample. Given that no companions were detected, it follows from binomial statistics that for a certain intrisic frequency $f$ of disk instability companions in the sample, the probability $P$ that such a frequency is consistent with our observations is given by:

\begin{equation}
P = \displaystyle \prod_i 1 - f p_i
\end{equation}

where $p_i$ is the detection probability of a given star in the sample. We use this procedure for all four combinations of cases (circular/elliptic and uniform/minimum) for the 15 B2--A0 stars in the sample. It follows that less than 29.0--31.8~\% of such stars form planets, brown dwarfs or very low-mass stars of $<$100~$M_{\rm jup}$ within 300~AU through disk instability, at 99~\% confidence. Here, 29.0~\% corresponds to the circular, uniform case and 31.8~\% to the eccentric, minimum case. Conversely, at 95~\% confidence, less than 19.9--21.8~\% form such objects and at 75~\% confidence, less than 9.7--10.6~\% do.

Individual notes for some targets follow:

\textbf{41 Ari:} 
There are no spectroscopically determined values for $T_{\rm eff}$ and $\log g$ for this target in the literature, so for the purpose of isochronal dating, we use values for these quantities determined through photometry instead \citep{allende1999}.

\textbf{Algol:} 
This star is the template system for Algol-type interacting binaries, where mass from a secondary giant star overflows its Roche lobe and transfers to the primary. In addition, there is a tertiary (non-interacting) component in the system \citep[e.g.][]{kim1989}. Due to the fact that mass transfer is taking place, it would not be applicable to use isochronal dating of the primary component, since such a procedure assumes a constant stellar mass over time. An accurate age estimation must account for the full evolutionary history of the system. Hence, we use an age estimation from \citet{sarna1993} that does precisely this. The best-fit age from that procedure, which we use here, is 460~Myr. We take the error on this quantity as the difference between the values of the two different models calculated in \citet{sarna1993}, $\sim$20 Myr. One candidate companion was found in the first-epoch data of Algol. By chance, Algol has a very small proper motion on the sky of 2.8~mas/yr. Hence, in contrast to all the other targets, the motion is too small for a statistically significant common proper motion test over the baseline of one or two years. However, since the projected separation of the candidate is about 145 AU and the total system mass is $\sim$6.2~$M_{\odot}$, one might expect some orbital motion if the candidate was real. For instance, at a circular face-on orbit, the motion would be 45~mas/yr. Furthermore, if it was a real companion it would be $\sim$13~$M_{\rm jup}$, and should be expected to have a significant non-zero value in CH4S-$H$, as opposed to a typical background star. Hence, we followed up the candidate in $H$-band. It shows no significant orbital motion, and with a contrast of 14.97$\pm$0.10 mag in $H$ compared to 15.10$\pm$0.11 mag in CH4S, it also shows no dimming at all relative to the star in $H$. We therefore conclude that it is very likely a background star.

\textbf{$\theta$ Hya:} 
While most stars can be dated isochronally to good precision, this is one of two targets that have relatively large uncertainties, since it has a relatively low mass and young age, where the isochrones are densely packed. The best-fitting isochrone is 130~Myr, but the fitting ages range from 60~Myr to 160~Myr. 

\textbf{$\beta$ Lib:} 
Like $\theta$ Hya, this is the other of the two cases where the isochronal dating has relatively large uncertainties, due to its relatively low mass and young age. The best-fitting isochrone is 80~Myr, but the fitting ages range from 40~Myr to 130~Myr. 

\textbf{$\gamma$~Crv:} 
The binary companion detected in our images has been previously discovered by \citet{roberts2007}, but without common proper motion confirmation. Comparing our data with the previous epoch, we can now confirm physical companionship: The secondary was located at 1.06$\pm$0.01\arcsec separation and 96.8$\pm$2.0$^{\rm o}$ position angle in 2002, and 1.119$\pm$0.006\arcsec and 106.5$\pm$0.1$^{\rm o}$ in 2009. This corresponds to a motion relative to the primary of 28~mas/yr, while the proper motion of the primary is 161~mas/yr, so a static background object can be excluded with high statistical significance. The orbital motion is also consistent with the magnitude of motion one might expect for a binary with these properties: A system with a total mass of 5~$M_{\odot}$ and a 50~AU orbit has an orbital period of 158 years. For a circular face-on orbit this would lead to a motion of 16$^{\rm o}$ over 7 years, which is of the same order as the 10$^{\rm o}$ motion that we measure. With contrasts of 7.1$\pm$0.1~mag in $I$-band and 5.10$\pm$0.14~mag in $K_{\rm c}$, isochrones \citep{marigo2008} suggest a secondary mass in the range of 0.8~$M_{\odot}$.

\textbf{109~Vir:} 
A likely binary companion was discovered to 109~Vir at 0.57\arcsec separation, 65.9$^{\rm o}$ position angle and a CH4S brightness contrast to the primary of 6.04~mag. It has not been confirmed through proper motion tests, but due to the small separation and contrast, it is unlikely to be an unrelated background source. If it is a real companion, isochrones \citep{marigo2008} suggest a secondary mass of approximately 0.65~$M_{\odot}$. 109~Vir is the only target in the B2--A0 sample with faint candidates that have not been followed up with proper motion tests. However, the two faint objects in the field would have too low masses to be able to form through disc instabilities (by a factor 6 each at their respective separations, and a factor 2--3 for anywhere else in the disk), so for the purpose of the analysis performed above, they can be stringently regarded as non-detections.

\textbf{Vega:} 
Since Vega is a quite nearby system at 7.8~pc, the NIRI field of view only reaches out to $\sim$90~AU. For this reason, we expand the field of view by including the contrast limits in a wider field of view provided in \citet{metchev2003} for H-band and \citet{macintosh2003} for K-band. This leads to that companions above approximately 7~$M_{\rm jup}$ can be excluded out to $\sim$230~AU.

\textbf{$\lambda$~Aql:} 
This star is located only 5$^{\rm o}$ from the galactic plane, 30$^{\rm o}$ from the center, and consequently the high-contrast images are crowded with many background stars. In total 105 objects have been registered. Since the two epochs had somewhat different orientations and one had better observing conditions leading to a better background-limited sensitivity, not all object that are visible or statistically significant in one epoch are likewise in the other. Here we focus on astrometry of those objects that are statistically significant in both epochs. This encompasses 55 objects, and makes us fully complete out to 10\arcsec and down to 17~mag contrast. Such a contrast corresponds to $\sim$4~$M_{\rm jup}$, which is more than a factor 2 lower than the lowest-mass objects that could form through gravitational instabilities in the disk of $\lambda$~Aql. The rest of the targets are considered unconfirmed.

\subsection{The White Dwarfs}
\label{s:wds}

For the two white dwarfs LHS~1870 and GD~140, total ages including both estimated main-sequence lifetime and post-main-sequence cooling time were adopted from \citet{burleigh2002}. Primordial and present-day masses were also adopted from that study. LHS~1870 is estimated to have had a main-sequence mass of 3.1~$M_{\odot}$, a white dwarf mass of 0.66~$M_{\odot}$ and an age of 700~Myr. GD~140 had an original mass of 6.3~$M_{\odot}$, a present-day mass of 0.9~$M_{\odot}$ and an age of 250~Myr. LHS~1870 has a modestly bright object in the field of view (see Appendix A). It has not been tested for proper motion, but if it were a real companion, it would have a mass of $\sim$15~$M_{\rm jup}$ at the system age, which would make it of order $\sim$30~\% as bright as LHS~1870 itself in M-band. This would make it fully detectable in the Spitzer excess study by \citet{mullally2007}, but their data show nothing to this effect. We therefore conclude that it is probably a background source.

Based on our mass sensitivity curves, we perform Monte-Carlo simulations in the same way as for the B-stars for the circular case, but without imposing any formation limits, in order to evaluate the general detection probabilies in mass versus semi-major axis space. During the mass loss phase of the white dwarfs, it can be expected that any planetary orbits will expand adiabatically, so that the semi-major axis increases by the same factor that the mass decreases with. This is a factor 4.7 in the case of LHS~1870 and a factor 7.0 in the case of GD~140. In other words, a planet residing at 2~AU during the main sequence phase of GD~140 would be expected to reside at 14~AU in the white dwarf phase. For the following analysis, we therefore divide the physical projected separation axis by the respective factor for the purpose of probing primordial semi-major axes in the system. 

\begin{figure}[ph]
\centering
\includegraphics[width=8cm]{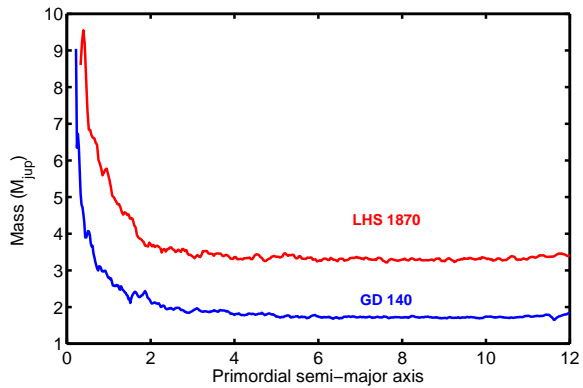}
\caption{\small{Detection limits for the white dwarfs GD 140 and LHS 1870, in the primordial separation range.}}
\label{f:wdlimits}
\end{figure}

\begin{figure}[ph]
\centering
\includegraphics[width=8cm]{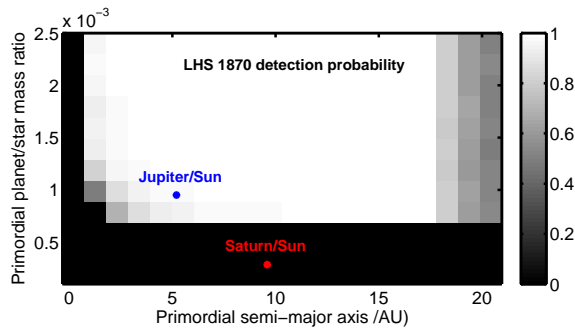}
\caption{\small{Detection probability map of LHS~1870, in primordial mass fraction and semi-major axis. A mass fraction equal to the Jupiter/Sun system at the relevant separation would be detectable.}}
\label{f:lhs1870}
\end{figure}

\begin{figure}[ph]
\centering
\includegraphics[width=8cm]{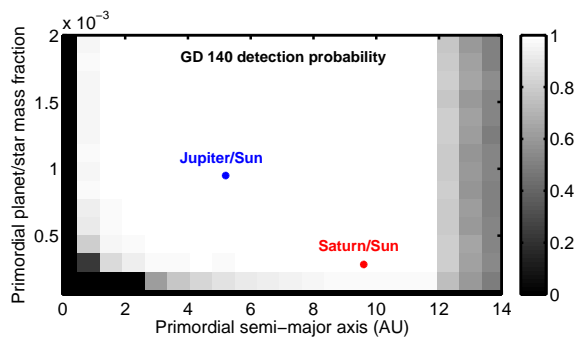}
\caption{\small{Detection probability map of GD~140, in primordial mass fraction and semi-major axis. Mass fractions equal to the Jupiter/Sun and Saturn/Sun systems at the relevant separations would be detectable.}}
\label{f:gd140}
\end{figure}

The mass detection limits of the two white dwarf targets are plotted in Fig \ref{f:wdlimits}. It is interesting to note that if planetary mass scales roughly linearly with stellar mass \citep[e.g.][]{alibert2011}, then the mass ratio between the star and planet may be a relevant quantity for what types of planets could be present in stellar systems of different masses, compared to our own. For this reason, we plot our detection probability spaces in primordial mass fraction versus primordial semi-major axis for LHS~1870 in Fig. \ref{f:lhs1870} and for GD~140 in Fig. \ref{f:gd140}, along with the Sun-Jupiter and Sun-Saturn values. In this sense, a Jupiter-equivalent planet would be detectable with near-unity probability in both systems (98~\% for LHS~1870, 100~\% for GD~140), and a Saturn-equivalent planet would be detectable with 97~\% probability for GD~140.

The radial velocity study of \citet{hekker2008} has implied that massive giant planets around massive stars (studied in their red giant phase) could be as much as 7 times more frequent than around Sun-like stars in the range of 1--2~AU (beyond this range, the study is incomplete). This is uncertain, since it is unclear to which extent false positives contribute to the frequency in the former case. If taken at face value however, and extrapolated to 20~AU, within which 17--20~\% of Sun-like stars are estimated to host giant planets \citep{cumming2008}, it would be expected that massive giant planets are essentially ubiquitous around massive stars. Another way to approach this issue is to extrapolate the planet freqeuency relation in \citet{johnson2010}, which predicts that giant planet frequency is directly proportional to stellar mass, in which case we might expect that giant planets should be ubiquitous from $\sim$5~$M_{\sun}$ and upwards, following the reasoning above. In the cases of LHS~1870 and GD~140, we probe down to 3.5~$M_{\rm jup}$ and 2~$M_{\rm jup}$ respectively, over essentially the full primordial separation range. Hence, down to these masses, it does not appear that planets are ubiquitous, although with only two targets studied to this high quality, it is presently not possible to quantify any meaningful upper limit. 

\subsection{Altair}
\label{s:altair}

In contrast to the B2--A0 stars, Altair is late-type enough that it is difficult to determine the age through temperature/gravity isochrones. Instead, we use the asteroseismological study of Altair by \citet{suarez2005}, which yields an age range of 250--750~Myr. We adopt 500~Myr as our baseline assumption for the age. 

\begin{figure}[ph]
\centering
\includegraphics[width=8cm]{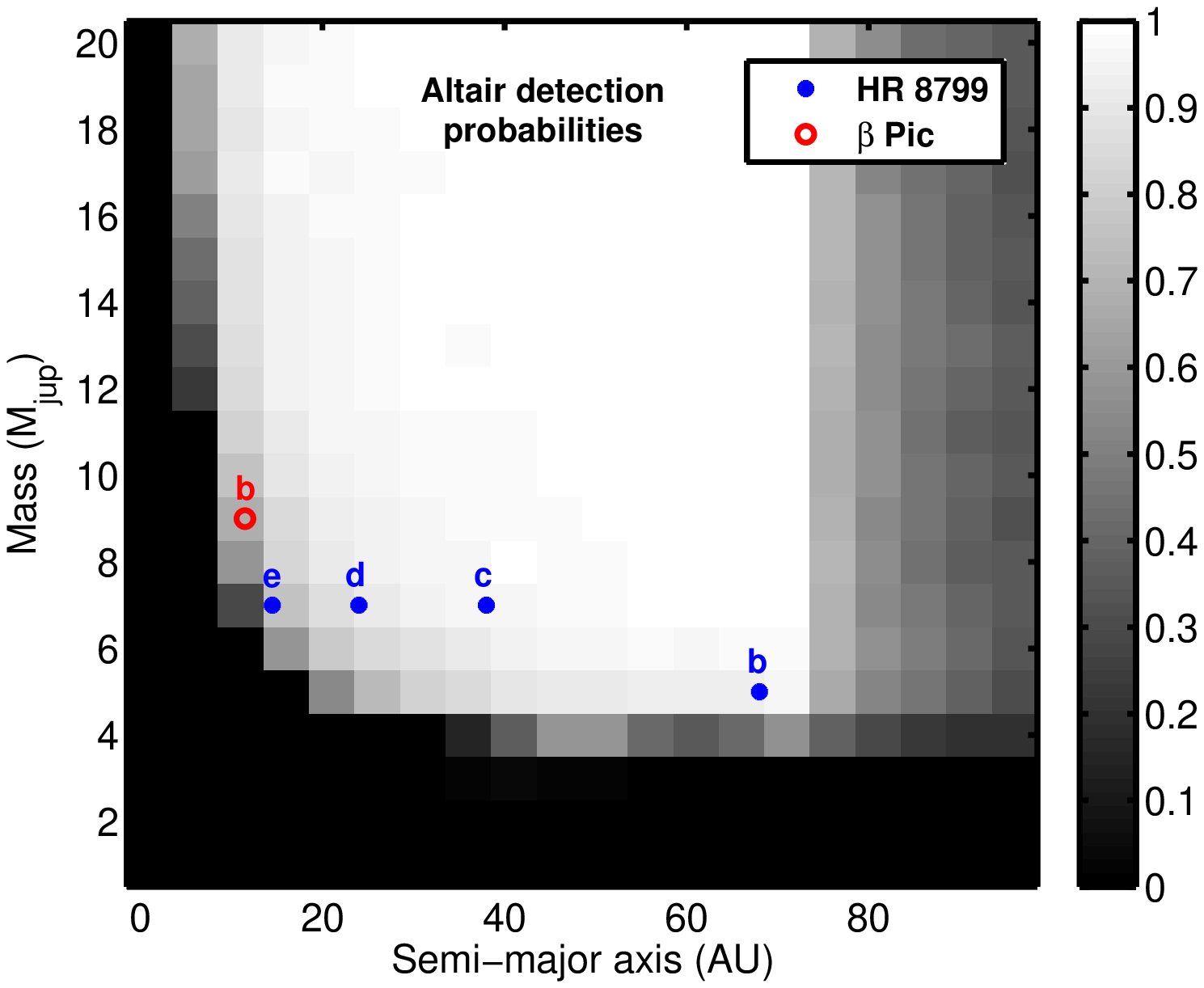}
\caption{\small{Detection probability map of Altair. Also plotted are the masses and separations of the $\beta$~Pic~b and HR~8799~b--e planets, all of which would be detectable in the image.}}
\label{f:altair}
\end{figure}

From the mass sensitivity curve at 500~Myr, we use Monte Carlo simulations in the same way as for the white dwarfs. The resulting mass versus semi-major axis detection space is plotted in Fig. \ref{f:altair}. Also plotted are the masses and physical projected separations of $\beta$~Pic~b \citep[][using the mean of the range 8--15~AU]{lagrange2010} and HR~8799~b--e \citep{marois2008,marois2010}. This provides an example of the utility of 4~$\mu$m data for nearby and relatively old stars -- despite the fact that Altair is more than an order of magnitude older than $\beta$~Pic and HR~8799, all the planets would still be detectable with this method if they had been present around Altair. The detection probabilities would be 68~\% for $\beta$~Pic~b and 94~\%, 96~\%, 92~\%, and 76~\% for HR~8799~b, c, d, and e, respectively. It is an interesting comparison also for the reason that the three stars are in the same spectral type range -- Altair has spectral type A7V, $\beta$ Pic A6V and HR 8799 A5V, according to SIMBAD.

\section{Discussion}
\label{s:discussion}

Since early on in the study of extrasolar planets, primarily two formation mechanisms have been discussed in the literature: Core accretion \citep[e.g.][]{pollack1996} and disk instabilitites \citep{boss2003}. Over the years, evidence has been building up that core accretion is the main formation mechanism for the population of close-in planets that can be studied with the radial velocity and transit techniques. For instance, the correlation between metallicity and giant planet frequency \citep[e.g.][]{fischer2005}, the fact that the giant planet radial velocity population can be broadly reproduced in population synthesis models based on core accretion \citep[e.g.][]{mordasini2009}, and the fact that the planet frequency increases with decreasing mass and radius \citep[e.g.][]{howard2010,borucki2011} all point to core accretion as the formation scenario. This does not by itself mean that core accretion is the dominant mechanism for the \textit{total} planet population. Instability-formed planets are expected only to be able to form far outside the separation range where radial velocity is presently sensitive. Hence, if disk instability would form and retain a larger number of planets in the far reaches of the system than core accretion manages closer in, then it would still be fair to say that disk instability would be the dominant mode of formation. In this paper, we explicitly address this question observationally, and find that around B-stars, which should in principle provide among the best possible places for disk instability formation of massive planets and brown dwarfs, such objects are rare. Previous surveys around Sun-like stars have not quantified limits with regard to the particular process of instabilities, but also reach the general conclusion that wide massive planets and brown dwarfs are rare \citep[e.g.][]{lafreniere2007b,chauvin2010,nielsen2010}. Meanwhile, estimations of planet frequency based on Kepler results indicate a planet frequency of $\sim$34~\% out to $\sim$0.5 AU \citep{borucki2011}, which is a tiny fraction of the separation range over which core accretion can form planets. Given that the frequency of giant planets increases for larger separations \citep[e.g.][]{cumming2008} and that planet frequency increases with decreasing mass in all domains where it has been properly tested \citep[e.g.][]{howard2010,borucki2011}, it seems inevitable that core accretion-formed planets will be essentially ubiquitous. Put together, these pieces of evidence enable us at this point to conclude with some confidence that core accretion is indeed the dominant planet formation mechanism.

This obviously does not necessarily mean that disk instabilities never form and retain planets. Indeed, as mentioned in the introduction it is the case that, although rare, massive planets do sometimes occur in orbits beyond where they could have been expected to form in-situ by core accretion \citep[e.g.][]{mordasini2009}. This does not constitute proof that they couldn't have formed in that way, because there are mechanisms by which a planet can increase its orbital separation, such as outward migration in a disk \citep[e.g.][]{crida2009} or planet-planet interactions \citep[e.g.][]{veras2009}, but it does mean that disk instability should also be considered in those cases. Potentially interesting companions in this regard exist both below the deuterium burning limit \citep[e.g.][]{kalas2005,lafreniere2008} and above it \citep[e.g.][]{biller2010,janson2011,lafreniere2011}. In many of these cases, regular binary formation is probably also an option -- although high-mass ratio systems do seem theoretically hard to produce that way \citep[e.g.][]{bate2005}, it is not yet clear whether this is true observationally for wide binaries \citep[e.g.][]{raghavan2010}. A particularly interesting planetary case is that of HR~8799 \citep{marois2008,marois2010}. HR~8799 is in many regards the most well-studied planetary system \citep[e.g.][]{janson2010,hinz2010,bowler2010,moro-martin2010,hinkley2011}, but its formation history remains mysterious. Planets 'b' and possibly 'c' are beyond the range where in-situ core accretion is expected to occur \citep[e.g.][]{mordasini2009,rafikov2011}. Meanwhile, planets 'd' and 'e' are inside of the range where instabilities could have occurred in-situ \citep[e.g.][]{rafikov2007}. Hence, regardless of how they formed, the planets must have undergone migration in one direction or the other. Although we cannot stringently exclude either scenario, we consider it of some significance that the imaged HR~8799 system is remarkably similar to the architecture of our own outer Solar system, after rescaling for equilibrium temperature, as shown in \citet{marois2010}. Following this comparison, it can be noted that Uranus and Neptune are also not expected to form in-situ in the same core accretion-based models \citep{mordasini2009}, yet few would doubt that this is the mechanism by which they formed \citep[see the Nice model for a discussion on the origin of the orbits of Uranus and Neptune, e.g.][]{tsiganis2005}.

A low frequency of planets formed and retained by disk instability does not mean that disk instabilities forming planetary fragments do not occur. Some studies have indicated that formed object in the gaseous disk may keep growing until they reach stellar masses \citep[e.g.][]{stamatellos2009,kratter2010}. Also, the timescale for instabilities is short and may place formation in a phase where the parent star is still in a cluster environment. In such environments, gravitational interactions between stellar systems are relatively common \citep[e.g.][]{malmberg2007}. Indeed, the fraction of wide binaries has been observationally shown to be likely affected by this mechanism in the dense Orion Nebular Cluster region \citep{reipurth2007}. Wide substellar companions with low binding energies may thus also undergo significant scattering.\footnote{Note that if such ejections occur, the objects become irrelevant to our study, since it concerns planets and brown dwarf companions, both of which are bound to the star by definition. For field objects in the corresponding mass range, such a mechanism may however be entirely relevant, this is not refuted or altogether addressed by our study.} Mutual interactions between multiple companions formed in the disk is another example of a mechanism that could cause scattering within the system or toward the field. Finally, one might imagine that planets and brown dwarfs which form far out in the disk migrate inwards through gaseous interaction, in the same way as expected for planets in the inner system. However, it would be difficult to hide substellar companions from detection in this manner. Firstly, such migration leads to substantial growth of the migrating object. \citet{mordasini2009} find that in the type II migration domain (which is what is relevant for disk instability-formed planets since they start out with large masses), growth occurs along lines of a fixed slope in log-log space, such that for every factor 2 that the planet moves inward, it gains a factor 8.8 in mass. If the disk runs out of gas for the planet to grow, the migration stops. Hence, for instance, if a 14~$M_{\rm jup}$ companion forms at 50 AU around the star Alpheratz in our sample (the lowest mass that can form at that separation for this star), and migrates into 25 AU, then its new mass is 123~$M_{\rm jup}$. This high mass means it is actually still detectable, but moreover, it is already well above the substellar boundary, and the limit for objects of interest to this study. Furthermore, even besides the growth issue, the rate of migration would have to be rather fine-tuned in order to hide the bulk of the gravitational instability population. Companions formed in the far regions of the disk have to move inward by well over a factor 5, hundreds of AU, in order to become potentially undetectable. Meanwhile, companions in the inner regions of the relevant space have to move inward by less than a factor 10, tens of AU, in order not to contaminate the inner planet population (thus diluting the brown dwarf desert and the metallicity correlation, etc.). For these reasons, we consider such migration unlikely. In any case, whether disk instability ever forms and retains planetary companions remains an open issue.

\section{Conclusions}

We have performed high-contrast imaging observations of 18 stars, of which 15 are nearby stars in the B2--A0 spectral type range, 2 are white dwarfs with massive progenitors, and one is the A7V star Altair. For the B2--A0 sub-sample, the resulting mass sensitivity was compared with predictions for planet and brown dwarf formation through disk instabilities, using Monte Carlo simulations for orbital distributions. For a range of conservative assumptions of where planets form within the boundaries allowed by disk instability, and a range of orbital distributions, it follows that $\sim$85~\% of massive planets, brown dwarfs, and very low-mass stars below 100~$M_{\rm jup}$ would be detectable within 300~AU on average. The null-detection among these targets therefore means that $<$32~\% of such stars form and retain in-situ companions through disk instability at 99~\% confidence, even though B-stars should be among the best possible places for such companions to form. Meanwhile, a range of results in the literature indicate that the frequency of planets formed by core accretion is $\gg$34~\% \citep[e.g.][]{borucki2011}. With these results taken together, there is now evidence that core accretion is likely the dominant formation mechanism for planetary companions to stars, and disk instability is not.

The two white dwarfs GD~140 and LHS~1870 provide a very high sensitivity to giant planets over the full primordial ranges where such objects could potentially be almost ubiquitous in the main-sequence phase. No companions were however detected. In the context of primordial star/planet mass fraction and semi-major axis, planets equivalent to the gas giants in our own system would have been detectable with high probability in the images, had they been present.

For the case of Altair, we demonstrated that planets equivalent to those in the HR~8799 and $\beta$~Pic systems would have been detectable in the system, despite the fact that Altair is more than an order of magnitude older. This is because the 4~$\mu$m technique used is well suited to the study of nearby and relatively old targets.

\acknowledgements
The authors thank Jonathan Fortney for providing thermal evolution tracks. We also thank the staff at the Gemini, Keck and Subaru telescopes for their help in performing these observations. The Gemini telescope is operated by the Association of Universities for Research in Astronomy, under a cooperative agreement with the NSF on behalf of the Gemini partnership. The Keck observatory is operated by the California Institute of Technology, the University of California and the National Aeronautics and Space Administration, and was made possible by a generous donation by the W.M. Keck foundation. The Keck time was allocated by NOAO. The Subaru telescope is operated by the National Astronomical Observatory of Japan. We acknowledge the cultural significance that the summit of Mauna Kea has to the indigenous Hawaiian community. M.J. is funded by the Reinhardt fellowship. R.J.'s research is supported by NSERC grants.

{\it Facilities:} \facility{Gemini:Gillett (NIRI)}, \facility{Keck:II (NIRC2), \facility{Subaru (IRCS)}}.

\appendix
\section{Point-sources and Contrast Curves}
Here we list all point sources that were detected in the high-contrast images (Table \ref{t:points}) and contrast curves for the various targets in different filters, at some specific separations (Table \ref{t:contrast}). The contrast values in Table \ref{t:points} refers to CH4S, unless stated otherwise.

\begin{longtable}{lcccccc}
\caption{Properties of all point sources.}\\
\hline\hline
HIP ID & Source no. & Sep. (\arcsec) & P.A. (deg) & $\Delta$ mag & Epoch & Status$^{\rm a}$ \\
\hline
\endfirsthead
\caption{continued.}\\
\hline\hline
HIP ID & Source no. & Sep. (\arcsec) & P.A. (deg) & $\Delta$ mag & Epoch & Status$^{\rm a}$ \\
\hline
\endhead
\hline
\endfoot
\hline
\multicolumn{7}{l}{\footnotesize{$^a$Status flag, where B = background source, C = confirmed companion, U = unconfirmed}}\\
\multicolumn{7}{l}{\footnotesize{$^b$Contrast in $H$-band.}}\\
\multicolumn{7}{l}{\footnotesize{$^c$Contrast in $K_{\rm c}$-band.}}\\
\hline
\endlastfoot
677	&	1	&	6.591$\pm$0.006	&	5.94$\pm$0.10	&	17.27$\pm$0.13	&	2008.79	&	B	\\
677	&	1	&	6.704$\pm$0.006	&	4.84$\pm$0.10	&	17.38$\pm$0.12	&	2009.58	&	B	\\
13209	&	1	&	9.051$\pm$0.006	&	229.06$\pm$0.10	&	17.02$\pm$0.19	&	2008.73	&	B	\\
13209	&	1	&	8.992$\pm$0.006	&	230.62$\pm$0.10	&	17.41$\pm$0.14	&	2010.64	&	B	\\
13209	&	2	&	7.328$\pm$0.006	&	33.90$\pm$0.10	&	15.73$\pm$0.11	&	2008.73	&	B	\\
13209	&	2	&	7.438$\pm$0.006	&	31.93$\pm$0.10	&	15.75$\pm$0.11	&	2010.64	&	B	\\
13209	&	3	&	12.222$\pm$0.006	&	25.93$\pm$0.10	&	12.92$\pm$0.10	&	2008.73	&	B	\\
13209	&	3	&	12.341$\pm$0.006	&	24.80$\pm$0.10	&	11.97$\pm$0.10	&	2010.64	&	B	\\
14576	&	1	&	5.089$\pm$0.006	&	206.49$\pm$0.10	&	15.11$\pm$0.11	&	2008.80	&	B	\\
14576	&	1	&	5.080$\pm$0.006	&	206.30$\pm$0.10	&	15.10$\pm$0.11	&	2009.74	&	B	\\
25336	&	1	&	5.781$\pm$0.006	&	8.28$\pm$0.10	&	17.57$\pm$0.16	&	2008.80	&	B	\\
25336	&	1	&	5.797$\pm$0.006	&	8.49$\pm$0.10	&	17.48$\pm$0.13	&	2010.76	&	B	\\
25336	&	2	&	12.323$\pm$0.006	&	52.11$\pm$0.10	&	17.68$\pm$0.17	&	2008.80	&	B	\\
25336	&	2	&	12.353$\pm$0.006	&	52.11$\pm$0.10	&	17.72$\pm$0.15	&	2010.76	&	B	\\
25336	&	3	&	10.381$\pm$0.006	&	246.36$\pm$0.10	&	16.94$\pm$0.12	&	2008.80	&	B	\\
25336	&	3	&	10.353$\pm$0.006	&	246.43$\pm$0.10	&	16.75$\pm$0.11	&	2010.76	&	B	\\
25428	&	1	&	8.366$\pm$0.006	&	357.08$\pm$0.10	&	16.90$\pm$0.11	&	2008.91	&	B	\\
25428	&	1	&	8.672$\pm$0.006	&	357.04$\pm$0.10	&	17.47$\pm$0.11	&	2010.77	&	B	\\
25428	&	2	&	11.776$\pm$0.006	&	0.78$\pm$0.10	&	16.86$\pm$0.11	&	2008.91	&	B	\\
25428	&	2	&	12.094$\pm$0.006	&	0.63$\pm$0.10	&	17.39$\pm$0.11	&	2010.77	&	B	\\
25428	&	3	&	11.607$\pm$0.006	&	295.75$\pm$0.10	&	13.69$\pm$0.10	&	2008.91	&	B	\\
25428	&	3	&	11.808$\pm$0.006	&	297.02$\pm$0.10	&	14.19$\pm$0.10	&	2010.77	&	B	\\
25428	&	4	&	11.795$\pm$0.006	&	297.07$\pm$0.10	&	14.29$\pm$0.10	&	2008.91	&	B	\\
25428	&	4	&	12.004$\pm$0.006	&	298.31$\pm$0.10	&	14.86$\pm$0.10	&	2010.77	&	B	\\
25428	&	5	&	9.824$\pm$0.006	&	69.79$\pm$0.10	&	16.07$\pm$0.11	&	2008.91	&	B	\\
25428	&	5	&	9.922$\pm$0.006	&	67.96$\pm$0.10	&	16.63$\pm$0.10	&	2010.77	&	B	\\
25428	&	6	&	10.856$\pm$0.006	&	79.54$\pm$0.10	&	18.16$\pm$0.16	&	2008.91	&	B	\\
25428	&	6	&	10.907$\pm$0.006	&	77.75$\pm$0.10	&	18.76$\pm$0.15	&	2010.77	&	B	\\
32560	&	1	&	3.736$\pm$0.002	&	46.40$\pm$0.02	&	5.70$\pm$0.20$^{\rm b}$	&	2009.96	&	U	\\
36188	&	1	&	10.472$\pm$0.006	&	352.15$\pm$0.10	&	14.47$\pm$0.10	&	2008.87	&	B	\\
36188	&	1	&	10.562$\pm$0.006	&	352.90$\pm$0.10	&	14.84$\pm$0.10	&	2010.91	&	B	\\
36188	&	2	&	7.541$\pm$0.006	&	204.66$\pm$0.10	&	16.55$\pm$0.11	&	2008.87	&	B	\\
36188	&	2	&	7.454$\pm$0.006	&	204.17$\pm$0.10	&	16.88$\pm$0.11	&	2010.91	&	B	\\
41307	&	1	&	5.820$\pm$0.006	&	86.68$\pm$0.10	&	15.62$\pm$0.11	&	2008.87	&	B	\\
41307	&	1	&	5.938$\pm$0.006	&	86.17$\pm$0.10	&	15.88$\pm$0.11	&	2010.91	&	B	\\
41307	&	2	&	9.566$\pm$0.006	&	246.18$\pm$0.10	&	15.60$\pm$0.11	&	2008.87	&	B	\\
41307	&	2	&	9.412$\pm$0.006	&	246.04$\pm$0.10	&	15.72$\pm$0.11	&	2010.91	&	B	\\
41307	&	3	&	10.080$\pm$0.006	&	221.30$\pm$0.10	&	16.73$\pm$0.14	&	2008.87	&	B	\\
41307	&	3	&	9.965$\pm$0.006	&	220.72$\pm$0.10	&	16.89$\pm$0.15	&	2010.91	&	B	\\
59803	&	1	&	1.119$\pm$0.006	&	106.52$\pm$0.10	&	5.10$\pm$0.14$^{\rm c}$	&	2009.09	&	C	\\
72220	&	1	&	5.731$\pm$0.006	&	281.66$\pm$0.10	&	16.33$\pm$0.12	&	2010.52	&	U	\\
72220	&	2	&	9.262$\pm$0.006	&	269.45$\pm$0.10	&	14.83$\pm$0.10	&	2010.52	&	U	\\
72220	&	3	&	0.572$\pm$0.006	&	65.91$\pm$0.10	&	6.04$\pm$0.10	&	2010.52	&	U	\\
93805	&	1	&	12.542$\pm$0.006	&	306.40$\pm$0.10	&	18.11$\pm$0.20	&	2010.46	&	U	\\
93805	&	2	&	12.048$\pm$0.006	&	302.38$\pm$0.10	&	13.27$\pm$0.10	&	2008.65	&	B	\\
93805	&	2	&	12.114$\pm$0.006	&	302.97$\pm$0.10	&	13.21$\pm$0.10	&	2010.46	&	B	\\
93805	&	3	&	10.784$\pm$0.006	&	290.45$\pm$0.10	&	17.64$\pm$0.13	&	2010.46	&	U	\\
93805	&	4	&	9.703$\pm$0.006	&	279.84$\pm$0.10	&	15.20$\pm$0.11	&	2008.65	&	B	\\
93805	&	4	&	9.714$\pm$0.006	&	280.68$\pm$0.10	&	14.92$\pm$0.10	&	2010.46	&	B	\\
93805	&	5	&	10.290$\pm$0.006	&	270.17$\pm$0.10	&	17.03$\pm$0.11	&	2010.46	&	U	\\
93805	&	6	&	9.484$\pm$0.006	&	270.83$\pm$0.10	&	18.17$\pm$0.18	&	2010.46	&	U	\\
93805	&	7	&	12.679$\pm$0.006	&	318.19$\pm$0.10	&	17.38$\pm$0.14	&	2010.46	&	U	\\
93805	&	8	&	10.343$\pm$0.006	&	306.60$\pm$0.10	&	17.88$\pm$0.14	&	2010.46	&	U	\\
93805	&	9	&	8.435$\pm$0.006	&	283.29$\pm$0.10	&	18.31$\pm$0.20	&	2010.46	&	U	\\
93805	&	10	&	11.141$\pm$0.006	&	318.11$\pm$0.10	&	17.80$\pm$0.14	&	2010.46	&	U	\\
93805	&	11	&	9.748$\pm$0.006	&	311.08$\pm$0.10	&	14.20$\pm$0.10	&	2008.65	&	B	\\
93805	&	11	&	9.836$\pm$0.006	&	311.77$\pm$0.10	&	14.02$\pm$0.10	&	2010.46	&	B	\\
93805	&	12	&	7.536$\pm$0.006	&	285.49$\pm$0.10	&	16.69$\pm$0.16	&	2008.65	&	B	\\
93805	&	12	&	7.575$\pm$0.006	&	286.63$\pm$0.10	&	16.52$\pm$0.11	&	2010.46	&	B	\\
93805	&	13	&	7.091$\pm$0.006	&	282.88$\pm$0.10	&	15.56$\pm$0.11	&	2008.65	&	B	\\
93805	&	13	&	7.115$\pm$0.006	&	284.03$\pm$0.10	&	15.57$\pm$0.10	&	2010.46	&	B	\\
93805	&	14	&	11.538$\pm$0.006	&	324.38$\pm$0.10	&	18.13$\pm$0.18	&	2010.46	&	U	\\
93805	&	15	&	10.414$\pm$0.006	&	324.56$\pm$0.10	&	15.24$\pm$0.11	&	2008.65	&	B	\\
93805	&	15	&	10.520$\pm$0.006	&	325.08$\pm$0.10	&	15.10$\pm$0.10	&	2010.46	&	B	\\
93805	&	16	&	9.900$\pm$0.006	&	320.31$\pm$0.10	&	17.01$\pm$0.11	&	2010.46	&	U	\\
93805	&	17	&	9.226$\pm$0.006	&	315.22$\pm$0.10	&	16.15$\pm$0.12	&	2008.65	&	B	\\
93805	&	17	&	9.320$\pm$0.006	&	315.91$\pm$0.10	&	16.12$\pm$0.11	&	2010.46	&	B	\\
93805	&	18	&	8.228$\pm$0.006	&	326.77$\pm$0.10	&	14.24$\pm$0.10	&	2008.65	&	B	\\
93805	&	18	&	8.340$\pm$0.006	&	327.38$\pm$0.10	&	14.32$\pm$0.10	&	2010.46	&	B	\\
93805	&	19	&	7.590$\pm$0.006	&	320.94$\pm$0.10	&	17.63$\pm$0.14	&	2010.46	&	U	\\
93805	&	20	&	7.845$\pm$0.006	&	316.40$\pm$0.10	&	18.17$\pm$0.19	&	2010.46	&	U	\\
93805	&	21	&	7.783$\pm$0.006	&	315.21$\pm$0.10	&	17.50$\pm$0.13	&	2010.46	&	U	\\
93805	&	22	&	6.564$\pm$0.006	&	310.48$\pm$0.10	&	17.67$\pm$0.16	&	2010.46	&	U	\\
93805	&	23	&	6.309$\pm$0.006	&	321.23$\pm$0.10	&	17.64$\pm$0.16	&	2010.46	&	U	\\
93805	&	24	&	5.500$\pm$0.006	&	316.10$\pm$0.10	&	17.18$\pm$0.14	&	2010.46	&	U	\\
93805	&	25	&	4.383$\pm$0.006	&	274.32$\pm$0.10	&	13.76$\pm$0.11	&	2008.65	&	B	\\
93805	&	25	&	4.384$\pm$0.006	&	276.26$\pm$0.10	&	13.97$\pm$0.10	&	2010.46	&	B	\\
93805	&	26	&	10.263$\pm$0.006	&	260.91$\pm$0.10	&	17.81$\pm$0.14	&	2010.46	&	U	\\
93805	&	27	&	8.957$\pm$0.006	&	253.58$\pm$0.10	&	17.69$\pm$0.14	&	2010.46	&	U	\\
93805	&	28	&	13.412$\pm$0.006	&	247.48$\pm$0.10	&	16.64$\pm$0.17	&	2008.65	&	B	\\
93805	&	28	&	13.344$\pm$0.006	&	248.02$\pm$0.10	&	16.05$\pm$0.11	&	2010.46	&	B	\\
93805	&	29	&	13.824$\pm$0.006	&	234.31$\pm$0.10	&	17.86$\pm$0.20	&	2010.46	&	U	\\
93805	&	30	&	11.537$\pm$0.006	&	236.38$\pm$0.10	&	15.97$\pm$0.12	&	2008.65	&	B	\\
93805	&	30	&	11.449$\pm$0.006	&	236.92$\pm$0.10	&	15.62$\pm$0.11	&	2010.46	&	B	\\
93805	&	31	&	11.103$\pm$0.006	&	231.42$\pm$0.10	&	17.25$\pm$0.12	&	2010.46	&	U	\\
93805	&	32	&	10.566$\pm$0.006	&	227.71$\pm$0.10	&	18.02$\pm$0.16	&	2010.46	&	U	\\
93805	&	33	&	9.238$\pm$0.006	&	228.83$\pm$0.10	&	18.20$\pm$0.17	&	2010.46	&	U	\\
93805	&	34	&	10.610$\pm$0.006	&	220.22$\pm$0.10	&	16.31$\pm$0.12	&	2008.65	&	B	\\
93805	&	34	&	10.508$\pm$0.006	&	220.72$\pm$0.10	&	16.13$\pm$0.11	&	2010.46	&	B	\\
93805	&	35	&	12.245$\pm$0.006	&	213.92$\pm$0.10	&	16.11$\pm$0.13	&	2008.65	&	B	\\
93805	&	35	&	12.127$\pm$0.006	&	214.31$\pm$0.10	&	16.10$\pm$0.11	&	2010.46	&	B	\\
93805	&	36	&	13.312$\pm$0.006	&	216.58$\pm$0.10	&	14.88$\pm$0.10	&	2010.46	&	U	\\
93805	&	37	&	8.350$\pm$0.006	&	228.17$\pm$0.10	&	12.51$\pm$0.10	&	2008.65	&	B	\\
93805	&	37	&	8.250$\pm$0.006	&	228.87$\pm$0.10	&	12.14$\pm$0.10	&	2010.46	&	B	\\
93805	&	38	&	6.115$\pm$0.006	&	228.11$\pm$0.10	&	17.09$\pm$0.13	&	2010.46	&	U	\\
93805	&	39	&	5.980$\pm$0.006	&	220.52$\pm$0.10	&	17.95$\pm$0.20	&	2010.46	&	U	\\
93805	&	40	&	5.900$\pm$0.006	&	216.69$\pm$0.10	&	17.40$\pm$0.15	&	2010.46	&	U	\\
93805	&	41	&	8.626$\pm$0.006	&	206.83$\pm$0.10	&	18.26$\pm$0.20	&	2010.46	&	U	\\
93805	&	42	&	4.422$\pm$0.006	&	228.15$\pm$0.10	&	16.20$\pm$0.18	&	2008.65	&	B	\\
93805	&	42	&	4.325$\pm$0.006	&	229.54$\pm$0.10	&	16.24$\pm$0.12	&	2010.46	&	B	\\
93805	&	43	&	4.277$\pm$0.006	&	228.62$\pm$0.10	&	16.58$\pm$0.27	&	2008.65	&	B	\\
93805	&	43	&	4.178$\pm$0.006	&	230.02$\pm$0.10	&	16.61$\pm$0.14	&	2010.46	&	B	\\
93805	&	44	&	8.073$\pm$0.006	&	343.95$\pm$0.10	&	17.45$\pm$0.13	&	2010.46	&	U	\\
93805	&	45	&	4.965$\pm$0.006	&	337.68$\pm$0.10	&	17.45$\pm$0.18	&	2010.46	&	U	\\
93805	&	46	&	4.730$\pm$0.006	&	338.41$\pm$0.10	&	17.78$\pm$0.23	&	2010.46	&	U	\\
93805	&	47	&	9.028$\pm$0.006	&	348.91$\pm$0.10	&	17.74$\pm$0.14	&	2010.46	&	U	\\
93805	&	48	&	9.869$\pm$0.006	&	354.96$\pm$0.10	&	16.74$\pm$0.14	&	2008.65	&	B	\\
93805	&	48	&	10.017$\pm$0.006	&	355.12$\pm$0.10	&	17.04$\pm$0.12	&	2010.46	&	B	\\
93805	&	49	&	7.644$\pm$0.006	&	359.63$\pm$0.10	&	18.01$\pm$0.18	&	2010.46	&	U	\\
93805	&	50	&	6.975$\pm$0.006	&	358.81$\pm$0.10	&	16.90$\pm$0.19	&	2008.65	&	B	\\
93805	&	50	&	7.122$\pm$0.006	&	358.94$\pm$0.10	&	17.71$\pm$0.15	&	2010.46	&	B	\\
93805	&	51	&	7.470$\pm$0.006	&	12.42$\pm$0.10	&	15.92$\pm$0.12	&	2008.65	&	B	\\
93805	&	51	&	7.618$\pm$0.006	&	12.20$\pm$0.10	&	16.19$\pm$0.11	&	2010.46	&	B	\\
93805	&	52	&	5.701$\pm$0.006	&	17.43$\pm$0.10	&	16.67$\pm$0.19	&	2008.65	&	B	\\
93805	&	52	&	5.851$\pm$0.006	&	17.02$\pm$0.10	&	17.11$\pm$0.13	&	2010.46	&	B	\\
93805	&	53	&	4.866$\pm$0.006	&	25.29$\pm$0.10	&	17.75$\pm$0.22	&	2010.46	&	U	\\
93805	&	54	&	9.667$\pm$0.006	&	17.56$\pm$0.10	&	16.27$\pm$0.12	&	2008.65	&	B	\\
93805	&	54	&	9.805$\pm$0.006	&	17.32$\pm$0.10	&	16.45$\pm$0.11	&	2010.46	&	B	\\
93805	&	55	&	5.248$\pm$0.006	&	37.53$\pm$0.10	&	14.82$\pm$0.11	&	2008.65	&	B	\\
93805	&	55	&	5.361$\pm$0.006	&	36.62$\pm$0.10	&	15.19$\pm$0.10	&	2010.46	&	B	\\
93805	&	56	&	4.841$\pm$0.006	&	47.47$\pm$0.10	&	13.34$\pm$0.10	&	2008.65	&	B	\\
93805	&	56	&	4.937$\pm$0.006	&	46.22$\pm$0.10	&	13.58$\pm$0.10	&	2010.46	&	B	\\
93805	&	57	&	4.256$\pm$0.006	&	69.57$\pm$0.10	&	15.08$\pm$0.13	&	2008.65	&	B	\\
93805	&	57	&	4.319$\pm$0.006	&	67.65$\pm$0.10	&	15.43$\pm$0.11	&	2010.46	&	B	\\
93805	&	58	&	3.503$\pm$0.006	&	92.55$\pm$0.10	&	13.21$\pm$0.11	&	2008.65	&	B	\\
93805	&	58	&	3.495$\pm$0.006	&	90.18$\pm$0.10	&	13.51$\pm$0.10	&	2010.46	&	B	\\
93805	&	59	&	4.719$\pm$0.006	&	124.92$\pm$0.10	&	16.16$\pm$0.15	&	2008.65	&	B	\\
93805	&	59	&	4.639$\pm$0.006	&	123.41$\pm$0.10	&	16.31$\pm$0.12	&	2010.46	&	B	\\
93805	&	60	&	6.493$\pm$0.006	&	50.51$\pm$0.10	&	17.65$\pm$0.16	&	2010.46	&	U	\\
93805	&	61	&	6.084$\pm$0.006	&	81.46$\pm$0.10	&	12.14$\pm$0.10	&	2008.65	&	B	\\
93805	&	61	&	6.100$\pm$0.006	&	80.09$\pm$0.10	&	12.27$\pm$0.10	&	2010.46	&	B	\\
93805	&	62	&	7.030$\pm$0.006	&	105.18$\pm$0.10	&	15.62$\pm$0.11	&	2008.65	&	B	\\
93805	&	62	&	6.983$\pm$0.006	&	104.07$\pm$0.10	&	15.68$\pm$0.11	&	2010.46	&	B	\\
93805	&	63	&	5.819$\pm$0.006	&	113.03$\pm$0.10	&	17.46$\pm$0.15	&	2010.46	&	U	\\
93805	&	64	&	4.734$\pm$0.006	&	204.71$\pm$0.10	&	15.89$\pm$0.20	&	2008.65	&	B	\\
93805	&	64	&	4.612$\pm$0.006	&	205.51$\pm$0.10	&	16.04$\pm$0.12	&	2010.46	&	B	\\
93805	&	65	&	7.426$\pm$0.006	&	196.74$\pm$0.10	&	16.03$\pm$0.19	&	2008.65	&	B	\\
93805	&	65	&	7.286$\pm$0.006	&	197.07$\pm$0.10	&	16.14$\pm$0.11	&	2010.46	&	B	\\
93805	&	66	&	4.038$\pm$0.006	&	170.35$\pm$0.10	&	13.62$\pm$0.11	&	2008.65	&	B	\\
93805	&	66	&	3.899$\pm$0.006	&	170.00$\pm$0.10	&	13.74$\pm$0.10	&	2010.46	&	B	\\
93805	&	67	&	6.772$\pm$0.006	&	173.61$\pm$0.10	&	14.08$\pm$0.10	&	2008.65	&	B	\\
93805	&	67	&	6.618$\pm$0.006	&	173.47$\pm$0.10	&	14.03$\pm$0.10	&	2010.46	&	B	\\
93805	&	68	&	9.306$\pm$0.006	&	172.00$\pm$0.10	&	14.62$\pm$0.10	&	2008.65	&	B	\\
93805	&	68	&	9.171$\pm$0.006	&	171.92$\pm$0.10	&	14.39$\pm$0.10	&	2010.46	&	B	\\
93805	&	69	&	7.070$\pm$0.006	&	163.80$\pm$0.10	&	17.68$\pm$0.15	&	2010.46	&	U	\\
93805	&	70	&	6.560$\pm$0.006	&	151.84$\pm$0.10	&	16.31$\pm$0.14	&	2008.65	&	B	\\
93805	&	70	&	6.443$\pm$0.006	&	151.34$\pm$0.10	&	16.49$\pm$0.11	&	2010.46	&	B	\\
93805	&	71	&	10.273$\pm$0.006	&	163.01$\pm$0.10	&	17.51$\pm$0.13	&	2010.46	&	U	\\
93805	&	72	&	11.424$\pm$0.006	&	165.61$\pm$0.10	&	16.97$\pm$0.12	&	2010.46	&	U	\\
93805	&	73	&	12.287$\pm$0.006	&	169.50$\pm$0.10	&	16.89$\pm$0.18	&	2008.65	&	B	\\
93805	&	73	&	12.167$\pm$0.006	&	169.45$\pm$0.10	&	16.45$\pm$0.11	&	2010.46	&	B	\\
93805	&	74	&	12.750$\pm$0.006	&	166.67$\pm$0.10	&	16.23$\pm$0.13	&	2008.65	&	B	\\
93805	&	74	&	12.609$\pm$0.006	&	166.56$\pm$0.10	&	15.37$\pm$0.10	&	2010.46	&	B	\\
93805	&	75	&	10.116$\pm$0.006	&	153.92$\pm$0.10	&	17.46$\pm$0.13	&	2010.46	&	U	\\
93805	&	76	&	11.804$\pm$0.006	&	155.30$\pm$0.10	&	17.34$\pm$0.13	&	2010.46	&	U	\\
93805	&	77	&	12.307$\pm$0.006	&	148.33$\pm$0.10	&	16.73$\pm$0.11	&	2010.46	&	U	\\
93805	&	78	&	7.524$\pm$0.006	&	115.28$\pm$0.10	&	13.64$\pm$0.10	&	2008.65	&	B	\\
93805	&	78	&	7.456$\pm$0.006	&	114.27$\pm$0.10	&	13.64$\pm$0.10	&	2010.46	&	B	\\
93805	&	79	&	8.442$\pm$0.006	&	123.27$\pm$0.10	&	16.34$\pm$0.13	&	2008.65	&	B	\\
93805	&	79	&	8.362$\pm$0.006	&	122.58$\pm$0.10	&	16.38$\pm$0.11	&	2010.46	&	B	\\
93805	&	80	&	8.481$\pm$0.006	&	130.64$\pm$0.10	&	17.50$\pm$0.13	&	2010.46	&	U	\\
93805	&	81	&	9.865$\pm$0.006	&	135.27$\pm$0.10	&	16.32$\pm$0.12	&	2008.65	&	B	\\
93805	&	81	&	9.760$\pm$0.006	&	134.75$\pm$0.10	&	16.30$\pm$0.11	&	2010.46	&	B	\\
93805	&	82	&	7.306$\pm$0.006	&	63.82$\pm$0.10	&	13.77$\pm$0.10	&	2008.65	&	B	\\
93805	&	82	&	7.369$\pm$0.006	&	62.85$\pm$0.10	&	14.13$\pm$0.10	&	2010.46	&	B	\\
93805	&	83	&	8.060$\pm$0.006	&	62.97$\pm$0.10	&	15.91$\pm$0.12	&	2008.65	&	B	\\
93805	&	83	&	8.116$\pm$0.006	&	62.13$\pm$0.10	&	16.24$\pm$0.11	&	2010.46	&	B	\\
93805	&	84	&	7.274$\pm$0.006	&	84.27$\pm$0.10	&	16.47$\pm$0.14	&	2008.65	&	B	\\
93805	&	84	&	7.284$\pm$0.006	&	83.05$\pm$0.10	&	16.71$\pm$0.11	&	2010.46	&	B	\\
93805	&	85	&	7.829$\pm$0.006	&	97.88$\pm$0.10	&	16.41$\pm$0.13	&	2008.65	&	B	\\
93805	&	85	&	7.821$\pm$0.006	&	96.86$\pm$0.10	&	16.64$\pm$0.11	&	2010.46	&	B	\\
93805	&	86	&	7.157$\pm$0.006	&	100.58$\pm$0.10	&	17.89$\pm$0.16	&	2010.46	&	U	\\
93805	&	87	&	8.520$\pm$0.006	&	72.48$\pm$0.10	&	12.94$\pm$0.10	&	2008.65	&	B	\\
93805	&	87	&	8.562$\pm$0.006	&	71.49$\pm$0.10	&	13.02$\pm$0.10	&	2010.46	&	B	\\
93805	&	88	&	8.421$\pm$0.006	&	78.83$\pm$0.10	&	17.55$\pm$0.14	&	2010.46	&	U	\\
93805	&	89	&	8.288$\pm$0.006	&	91.88$\pm$0.10	&	18.25$\pm$0.10	&	2010.46	&	U	\\
93805	&	90	&	8.286$\pm$0.006	&	94.21$\pm$0.10	&	18.15$\pm$0.20	&	2010.46	&	U	\\
93805	&	91	&	9.042$\pm$0.006	&	83.03$\pm$0.10	&	16.01$\pm$0.12	&	2008.65	&	B	\\
93805	&	91	&	9.051$\pm$0.006	&	82.16$\pm$0.10	&	16.26$\pm$0.11	&	2010.46	&	B	\\
93805	&	92	&	10.085$\pm$0.006	&	98.78$\pm$0.10	&	14.81$\pm$0.10	&	2008.65	&	B	\\
93805	&	92	&	10.055$\pm$0.006	&	97.98$\pm$0.10	&	14.79$\pm$0.10	&	2010.46	&	B	\\
93805	&	93	&	11.455$\pm$0.006	&	84.47$\pm$0.10	&	16.46$\pm$0.13	&	2008.65	&	B	\\
93805	&	93	&	11.455$\pm$0.006	&	83.72$\pm$0.10	&	16.20$\pm$0.10	&	2010.46	&	B	\\
93805	&	94	&	12.316$\pm$0.006	&	67.33$\pm$0.10	&	17.09$\pm$0.12	&	2010.46	&	U	\\
93805	&	95	&	12.014$\pm$0.006	&	64.24$\pm$0.10	&	14.02$\pm$0.10	&	2008.65	&	B	\\
93805	&	95	&	12.071$\pm$0.006	&	63.63$\pm$0.10	&	14.01$\pm$0.10	&	2010.46	&	B	\\
93805	&	96	&	13.854$\pm$0.006	&	53.75$\pm$0.10	&	16.63$\pm$0.16	&	2008.65	&	B	\\
93805	&	96	&	13.943$\pm$0.006	&	53.25$\pm$0.10	&	16.81$\pm$0.12	&	2010.46	&	B	\\
93805	&	97	&	14.130$\pm$0.006	&	50.72$\pm$0.10	&	16.03$\pm$0.13	&	2008.65	&	B	\\
93805	&	97	&	14.214$\pm$0.006	&	50.32$\pm$0.10	&	16.34$\pm$0.11	&	2010.46	&	B	\\
93805	&	98	&	13.738$\pm$0.006	&	47.88$\pm$0.10	&	16.58$\pm$0.15	&	2008.65	&	B	\\
93805	&	98	&	13.836$\pm$0.006	&	47.39$\pm$0.10	&	16.87$\pm$0.12	&	2010.46	&	B	\\
93805	&	99	&	11.381$\pm$0.006	&	135.53$\pm$0.10	&	17.26$\pm$0.18	&	2008.65	&	B	\\
93805	&	99	&	11.270$\pm$0.006	&	135.09$\pm$0.10	&	16.88$\pm$0.11	&	2010.46	&	B	\\
93805	&	100	&	12.784$\pm$0.006	&	140.37$\pm$0.10	&	17.22$\pm$0.13	&	2010.46	&	U	\\
93805	&	101	&	11.815$\pm$0.006	&	131.91$\pm$0.10	&	16.54$\pm$0.14	&	2008.65	&	B	\\
93805	&	101	&	11.718$\pm$0.006	&	131.37$\pm$0.10	&	16.44$\pm$0.11	&	2010.46	&	B	\\
93805	&	102	&	14.252$\pm$0.006	&	132.85$\pm$0.10	&	17.20$\pm$0.14	&	2010.46	&	U	\\
93805	&	103	&	14.679$\pm$0.006	&	136.56$\pm$0.10	&	17.32$\pm$0.14	&	2010.46	&	U	\\
93805	&	104	&	12.775$\pm$0.006	&	348.23$\pm$0.10	&	15.62$\pm$0.11	&	2008.65	&	B	\\
93805	&	104	&	12.919$\pm$0.006	&	348.38$\pm$0.10	&	14.95$\pm$0.10	&	2010.46	&	B	\\
93805	&	105	&	7.072$\pm$0.006	&	236.74$\pm$0.10	&	14.93$\pm$0.11	&	2008.65	&	B	\\
93805	&	105	&	6.990$\pm$0.006	&	237.74$\pm$0.10	&	14.84$\pm$0.10	&	2010.46	&	B	\\
\label{t:points}
\end{longtable}

\begin{table*}[p]
\caption{Contrast limits of the survey (5$\sigma$)}
\label{t:contrast}
\centering

\begin{tabular}{lcccccccccc}
\hline
\hline
HIP ID	&	Filt.	&	0.5\arcsec	&	0.75\arcsec	&	1.0\arcsec	&	2.0\arcsec	&	3.0\arcsec	&	4.0\arcsec	&	5.0\arcsec	&	7.5\arcsec	&	10.0\arcsec	\\
\hline
677	&	CH4S	&	...	&	...	&	...	&	...	&	17.1	&	18.0	&	18.7	&	19.6	&	20.1	\\
677	&	Br$\alpha$	&	7.5	&	8.3	&	8.7	&	9.5	&	9.8	&	10.0	&	10.1	&	10.2	&	10.2	\\
13209	&	CH4S	&	...	&	...	&	...	&	15.2	&	16.6	&	17.3	&	17.9	&	18.4	&	18.5	\\
13209	&	Br$\alpha$	&	8.4	&	8.4	&	8.6	&	8.8	&	9.0	&	8.9	&	9.0	&	9.0	&	9.0	\\
14576	&	CH4S	&	...	&	...	&	...	&	...	&	15.1	&	16.2	&	16.8	&	18.0	&	18.6	\\
14576	&	Br$\alpha$	&	9.3	&	9.8	&	9.9	&	10.2	&	10.5	&	10.5	&	10.5	&	10.5	&	10.5	\\
25336	&	CH4S	&	...	&	...	&	...	&	...	&	16.6	&	17.6	&	18.2	&	19.2	&	19.6	\\
25336	&	$K_{\rm c}$	&	...	&	...	&	13.4	&	15.5	&	16.6	&	17.3	&	18.1	&	19.0	&	19.4	\\
25428	&	CH4S	&	...	&	...	&	...	&	...	&	16.7	&	17.5	&	18.0	&	19.3	&	19.8	\\
25428	&	$K_{\rm c}$	&	...	&	...	&	13.7	&	15.4	&	16.9	&	17.3	&	18.2	&	19.2	&	19.6	\\
32560	&	$H$	&	10.1	&	10.9	&	11.0	&	11.4	&	11.3	&	11.3	&	11.3	&	...	&	...	\\
36188	&	CH4S	&	...	&	...	&	...	&	15.0	&	15.9	&	16.9	&	17.7	&	18.7	&	19.0	\\
36188	&	$K_{\rm c}$	&	...	&	11.9	&	12.6	&	14.8	&	16.1	&	16.9	&	17.5	&	18.4	&	18.7	\\
41307	&	CH4S	&	...	&	...	&	...	&	15.0	&	16.1	&	16.9	&	17.5	&	18.0	&	18.2	\\
41307	&	$K_{\rm c}$	&	...	&	11.7	&	12.8	&	14.6	&	16.0	&	16.8	&	17.4	&	18.0	&	18.1	\\
45336	&	CH4S	&	...	&	...	&	...	&	14.0	&	15.1	&	16.1	&	16.3	&	17.2	&	17.5	\\
45336	&	$K_{\rm c}$	&	...	&	11.6	&	12.7	&	14.7	&	15.9	&	16.6	&	17.0	&	17.5	&	17.5	\\
49669	&	CH4S	&	...	&	...	&	...	&	...	&	14.7	&	15.7	&	16.5	&	17.5	&	18.0	\\
49669	&	$K_{\rm c}$	&	11.8	&	12.7	&	13.5	&	15.5	&	16.0	&	...	&	...	&	...	&	...	\\
56662	&	$H$	&	9.2	&	9.5	&	10.1	&	11.0	&	11.3	&	11.3	&	11.2	&	...	&	...	\\
59803	&	$K_{\rm c}$	&	...	&	13.1	&	13.9	&	16.1	&	17.3	&	17.9	&	18.6	&	19.2	&	19.4	\\
72220	&	CH4S	&	9.7	&	11.3	&	12.0	&	14.6	&	15.9	&	17.0	&	17.5	&	18.2	&	18.4	\\
74785	&	CH4S	&	10.2	&	11.6	&	12.4	&	14.4	&	15.8	&	17.1	&	17.7	&	18.6	&	19.0	\\
83207	&	CH4S	&	11.0	&	12.3	&	13.2	&	15.2	&	16.7	&	17.4	&	18.1	&	18.8	&	19.1	\\
91262	&	Br$\alpha$	&	11.1	&	11.6	&	12.1	&	12.9	&	12.9	&	13.0	&	13.0	&	13.0	&	13.2	\\
93805	&	CH4S	&	...	&	...	&	...	&	15.4	&	16.7	&	17.4	&	17.8	&	18.5	&	18.6	\\
93805	&	Br$\alpha$	&	7.5	&	7.8	&	8.6	&	9.4	&	9.5	&	9.6	&	9.6	&	9.8	&	9.8	\\
97649	&	Br$\alpha$	&	8.8	&	10.5	&	11.1	&	13.0	&	13.9	&	14.4	&	14.5	&	14.5	&	14.5	\\
\hline
\end{tabular}
\end{table*}

\clearpage

\end{document}